\documentclass[
 aps,
 jmp,%
 amsmath,amssymb,
preprint,%
]{revtex4-2}

\usepackage{graphicx}
\usepackage{dcolumn}
\usepackage{bm}
\usepackage{multirow}
\usepackage{listings} 
\usepackage[english]{babel} 
\begin{document}


\title[A Superconducting Nanowire-based Architecture for Neuromorphic Computing]{A Superconducting Nanowire-based Architecture for Neuromorphic Computing}

\author{Andres E. Lombo}
\affiliation{Division of Engineering Science, University of Toronto, Toronto, Canada}

\author{Jesus Lares}
\affiliation{Electrical Engineering and Computer Science Department, Massachusetts Institute of Technology, Cambridge, United States of America}

\author{Matteo Castellani}
\affiliation{Electrical Engineering and Computer Science Department, Massachusetts Institute of Technology, Cambridge, United States of America}
\author{Chi-Ning Chou}
\affiliation{School of Engineering and Applied Sciences, Harvard University, Cambridge, United States of America}
\author{Nancy Lynch}
\affiliation{Electrical Engineering and Computer Science Department, Massachusetts Institute of Technology, Cambridge, United States of America}
\author{Karl K. Berggren}
\affiliation{Electrical Engineering and Computer Science Department, Massachusetts Institute of Technology, Cambridge, United States of America}


\begin{abstract}
Neuromorphic computing would benefit from the utilization of improved customized hardware. However, the translation of neuromorphic algorithms to hardware is not easily accomplished.  In particular, building superconducting neuromorphic systems requires expertise in both superconducting physics and theoretical neuroscience, which makes such design particularly challenging. In this work, we aim to bridge this gap by presenting a tool and methodology to translate algorithmic parameters into circuit specifications. We first show the correspondence between theoretical neuroscience models and the dynamics of our circuit topologies. We then apply this tool to solve a linear system and implement Boolean logic gates by creating spiking neural networks with our superconducting nanowire-based hardware. 
\end{abstract}

\maketitle

\section{Introduction}
\noindent
Neuromorphic computing attempts to mimic the behavior of biological neurons and synapses in the human brain. Recently, increased understanding of the physics of devices for neuromorphic computing \cite{markovic_physics_2020,berggren_roadmap_2020}, and the theory of algorithms for neuromorphic computing \cite{lynch_basic_2021} has led to the development of CMOS-based neuromorphic architectures \cite{davies_loihi_2018-1} that are three orders of magnitude more efficient in terms of their energy-delay product when compared to traditional multiply-and-accumulate operations \cite{davies_lessons_2021}. However, these systems are also nowhere near the power figure of merit (energy per operation) required to create a massive scale neuromorphic computer, such as the human brain.

    For these reasons, efforts to mimic neurons and synapses may need to move towards  systems that have an intrinsic spiking ability and extremely low energy consumption. This is the case in superconducting electronics where the constituent elements exhibit nonlinear characteristics and have very low or no power dissipation. Superconducting circuits offer drastically lower power consumption even when cryogenic cooling energy costs is taken into account \cite{holmes_energy-efficient_2013, toomey_design_2019}. Previous developments of neuromorphic architectures using superconducting electronics have used Josephson junctions \cite{goteti_superconducting_2021,schneider_fan-out_2020,crotty_josephson_2010,segall_phase-flip_2014}, quantum-phase slip junctions \cite{cheng_superconducting_2019,cheng_spiking_2018,cheng_toward_2021}, magnetic tunnel junctions \cite{schneider_ultralow_2018}, systems with Josephson junctions and superconducting nanowire single photon detectors \cite{shainline_superconducting_2017,shainline_superconducting_2019}, and nanowires as relaxation oscillators \cite{toomey_design_2019} to construct circuits that emulate biological neurons and synapses. Superconducting nanowires offer ease of fabrication and are most easily integrated with classical circuit elements. In addition, the ability of superconducting circuits to operate with near-lossless interconnects makes them an attractive choice for implementing a low-power neuromorphic architecture.

Simultaneously, the success of Artificial Neural Networks (ANNs) \cite{krizhevsky_imagenet_2012} in computing applications such as pattern recognition and natural language processing coupled with the widespread adoption of machine learning methods in many areas of science and engineering is an indication that abstracting a computing problem and eliminating hardware dependencies is a promising approach for going beyond Moore’s law. Among these networks, Spiking Neural Networks (SNNs) closely simulate the dynamics of biological neurons and synapses in the brain. Approaches to spiking neural networks that possess brain-like properties have gained increased attention due to their widespread use in applications spanning decision making \cite{ye_quantifying_2021}, image recognition \cite{tapson_synthesis_2013} and optimization problems \cite{maass_energy-efficient_2016}.

The direct translation of superconducting neuromorphic architectures into algorithmic formulations of a problem has been little explored, and a complete description of an algorithmic implementation in neuromorphic hardware remains to be seen.  It is very difficult for hardware designers to condense an abstract algorithmic problem into a specific hardware platform without increasing the complexity of circuits or compromising energy efficiency.  This gap in the field stems from the difficulty in reconciling the algorithmic and hardware oriented approaches. At this point in time, this issue is often a question of expertise: hardware designers do not have ready knowledge of algorithmic subtleties, while algorithm developers do not have access to the hardware. Thus, there is a need for a tool to allow computer scientists to test algorithms on neuronal circuits. We devise and present such a tool here.

In this work, we address the issue of translation from theoretical algorithms to a specific implementation by using the example of solving linear systems with a superconducting neuromorphic network. We show the direct relation between a basic compositional model as well as a leaky integrate-and-fire model and the proposed superconducting nanowire-based neuromorphic hardware. We compile this correspondence into a tool to translate between hardware and algorithmic descriptions of the neuromorphic architecture. We conclude with a discussion on the outlook of the scaling of superconducting nanowire-based circuits in the context of neural networks. 

\section{Methods}
\noindent
The building blocks of the hardware architecture are superconducting nanowires \cite{berggren_superconducting_2018} and hTrons \cite{baghdadi_multilayered_2020}. In a superconducting nanowire biased with a current, superconductivity breaks down when the current when the current exceeds the critical current $I_c$. As a consequence, the nanowire develops a resistance $R_{hs}$ and a voltage $v=i_{nw}R_{hs}$. Superconductivity is restored in the nanowire when its current is reduced below the retrapping current $I_r$. When a superconducting nanowire is placed in parallel with a resistor, the relaxation from the normal to the superconducting state of the nanowire can couple with the resistor. When biased by a current above $I_c$, the nanowire switches and electrothermal feedback produces continuous voltage spikes across the nanowire. This is termed a relaxation oscillator \cite{toomey_frequency_2018}.

The hTron is a circuit element that acts as a thermally activated switch. It consists of a superconducting nanowire (the channel), placed in close proximity to a resistive element (the gate) \cite{baghdadi_multilayered_2020}. When the channel is biased by a current below its threshold $I_{c,h}$, heat dissipated by the gate can increase the temperature of the channel and break superconductivity. Superconductivity is restored in the channel when it has cooled and its current is reduced below a threshold $I_{r,h}$. This threshold is dependent on the temperature of the channel and decreases for increasing temperature \cite{baghdadi_multilayered_2020}. 

The simulations used in this work are based on models for superconducting nanowires \cite{berggren_superconducting_2018}  and for hTrons \cite{baghdadi_multilayered_2020} implemented in LTSPICE \cite{castellani_design_nodate}.

\subsection{Hardware Design}
\noindent
In previous work \cite{toomey_design_2019}, the application of two superconducting nanowires whose intrinsic nonlinear inductance $L_{nw}$ was used to generate spiking behaviour is presented. We summarize the description here.

\begin{figure*}
    \centering
    \includegraphics[width=\textwidth]{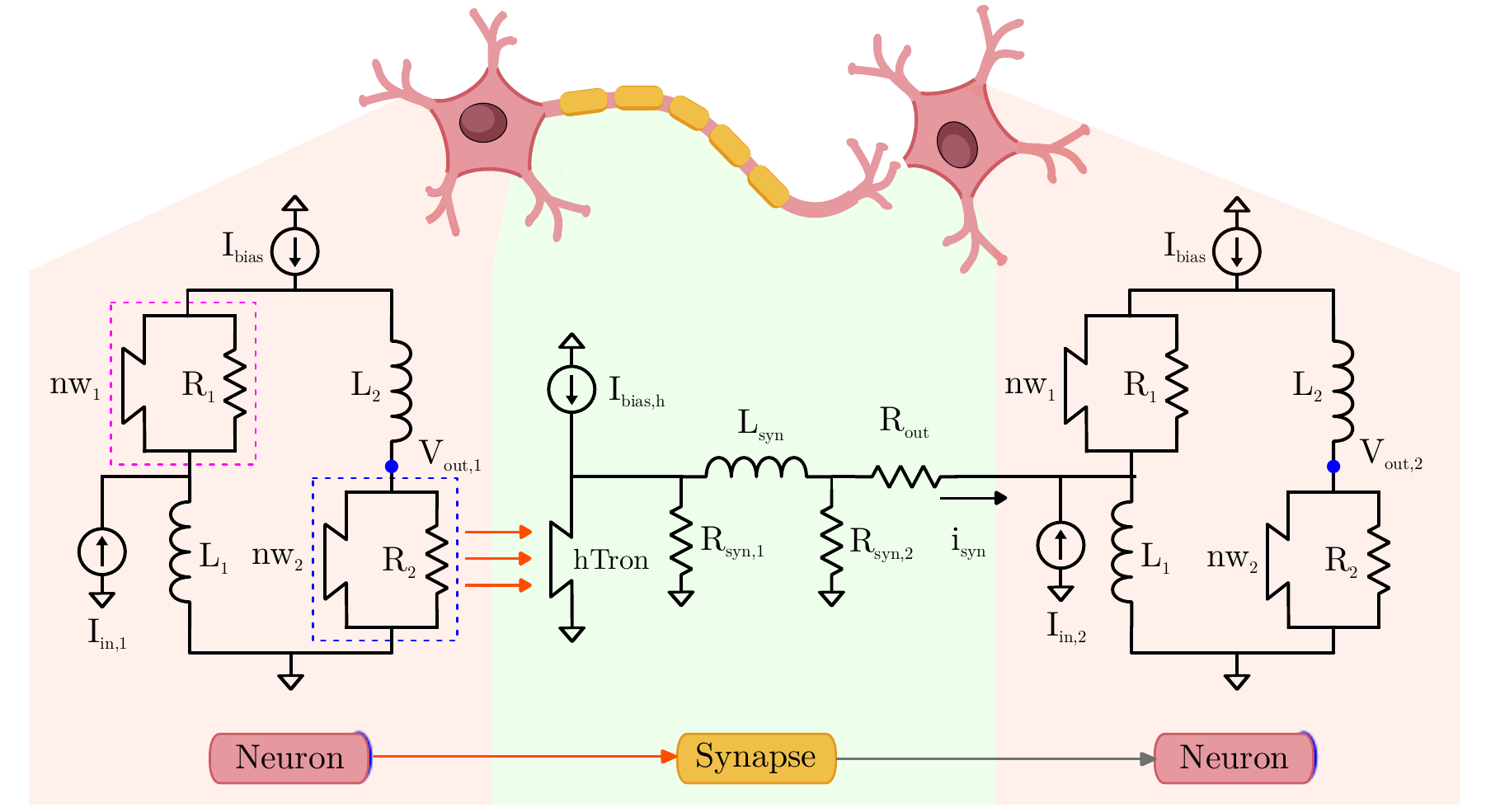}
    \caption{Circuit topology for a circuit consisting of an input neuron (left) upstream, a synapse (centre), and a target neuron (right) downstream. The neuron has a control (pink) and main (blue) relaxation oscillator each with a nanowire $nw_1$; $nw_2$. Voltage spikes at node $V_{out,1}$ generate heat (orange arrows) in close proximity of the hTron in the synapse which will be transferred to the target neuron via $R_{out}$.}
    \label{fig:1}
\end{figure*}

As illustrated in Figure \ref{fig:1}, the nanowire neuron consists of a main and control relaxation oscillator in a loop. A source $I_{bias}$ biases both oscillators below their critical currents but in opposite directions. An input current pulse at $I_{in}$ applied to the loop then results in the current in the main oscillator to exceed $I_c$, causing it to switch. Then, current is diverted counterclockwise in the loop which causes the control oscillator to switch while the main oscillator relaxes. The switching of the control oscillator diverts current clockwise and causes the main oscillator to switch again. Each time main oscillator switches, a voltage spike will be seen at node $V_{out}$. The main and control oscillator act analogously to the Na+ and K+ ion channels in the Hodgkin-Huxley neuron model \cite{hodgkin_quantitative_1952}.

The synapse consists of an hTron and an integration loop formed by $L_{syn}, R_{syn,1}$, and $R_{syn,2}$. When a voltage spike appears at $V_{out}$,  heat is dissipated across $R_2$ (orange arrows). $R_2$ acts as the gate of the hTron, and the heat from it lowers the critical current of the hTron channel, causing the hTron to switch. In the switched (normal, non-superconducting) state, the resistance of the channel is typically on the order of $10^2\; \Omega$ for NbN films. The typical resistance for $R_{syn,1}$ and $R_{syn,2}$ is on the order of $10\;\Omega$.  Thus, when the hTron channel switches, the majority of the current $I_{bias,h}$ is diverted into the integration loop. A portion of the current through $L_{syn}$ is then transmitted to the target neuron via $R_{out}$. This process is analogous to the integration behaviour of the Hodgkin-Huxley model \cite{hodgkin_quantitative_1952}.

\begin{figure}
    \centering
    \includegraphics{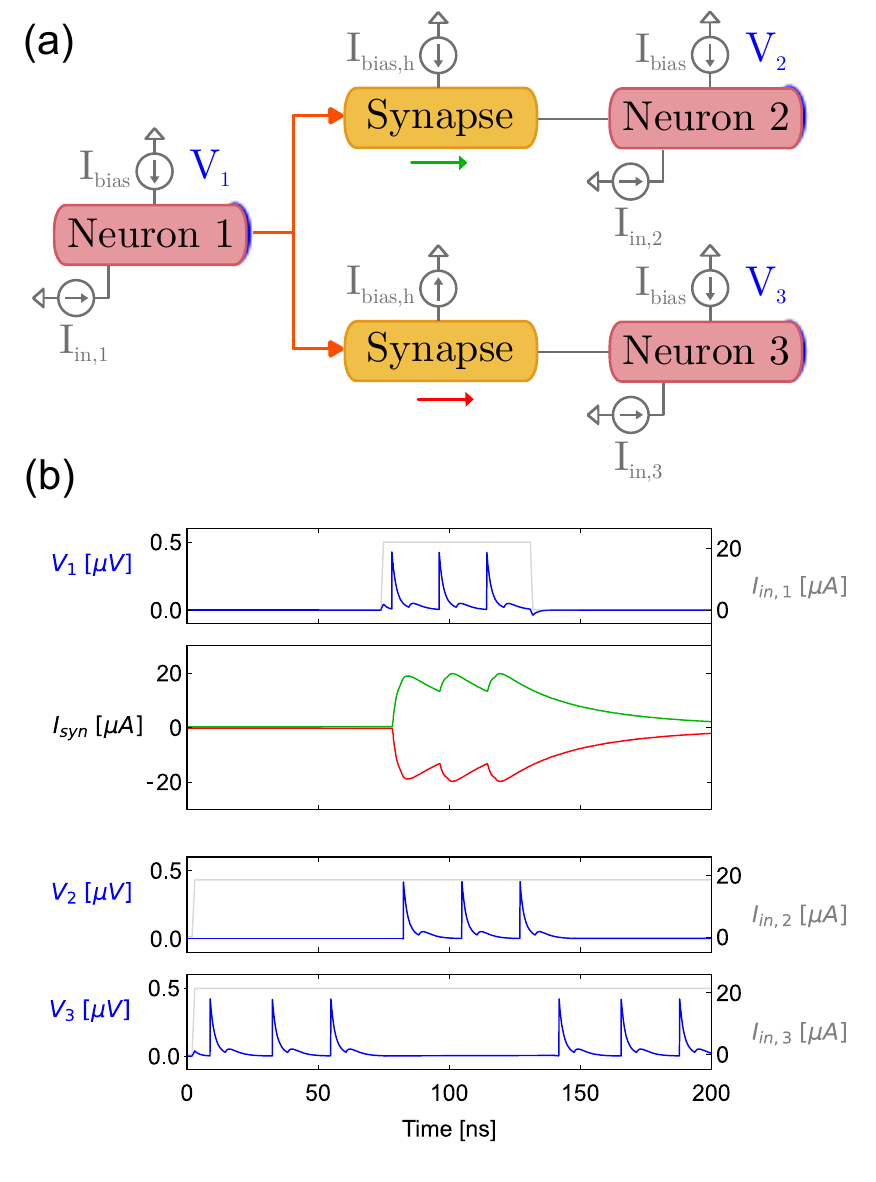}
    \caption{Simulation results of excitatory and inhibitory connections between neurons. (a) Circuit schematic for a neuron-synapse-neuron network (b) Waveforms showing the spikes $V_1$ of neuron 1, the current in the synapses $I_{syn}$, and the spikes $V_2$ and $V_3$ of neurons 2 and 3. In this simulation, $I_{in,1}=22\; \mu A, I_{in,2}=19\; \mu A $ and $ I_{in,3}=22\; \mu A$.}
    \label{fig:2}
\end{figure}

The simulated operation of a simple neuron-synapse-neuron connection demonstrating excitatory and inhibitory behavior is shown in figure \ref{fig:2}. The network illustrated consists of an input neuron (1) connected to two target neurons (2,3). The spike output from neuron 1 is connected thermally to the synapses via the hTron and the synapses are connected electrically to neuron (2,3).  When $I_{in,1}$ makes neuron 1 fire, the spikes are integrated in the synapse as shown by $I_{syn}$ in figure \ref{fig:2}b. This synaptic current leads to the excitation of neuron 2 for a brief period of time. Similarly, the same current $I_syn$ but in the opposite direction leads to inhibition of the firing of neuron 3 for a brief period of time.

\begin{figure*}
    \centering
    \includegraphics[width=\textwidth]{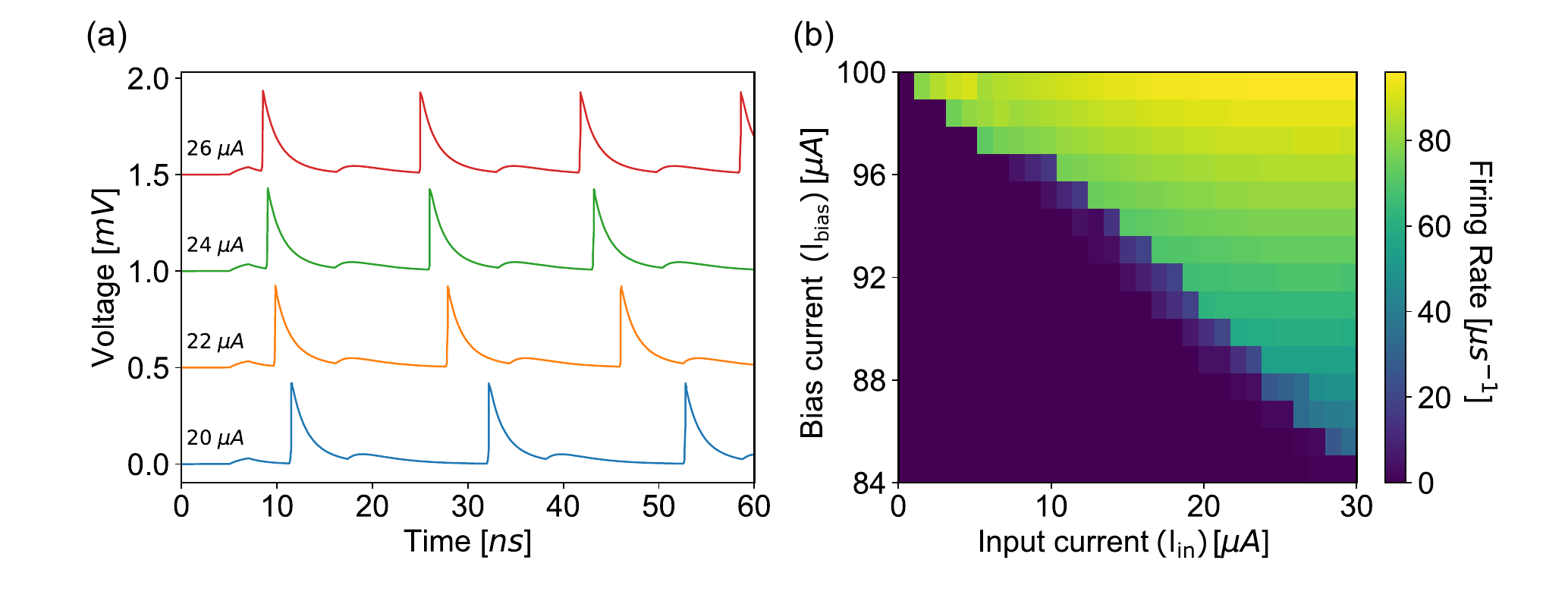}
    \caption{Simulation results of the frequency tunability of the nanowire neuron. (a) Plot of spike waveforms at neuron output with varying  $I_{in}$. Note that the waveforms are offset vertically for clarity. Spiking frequency increases with increased input current to the neuron. (b) Colour map of the firing rate (inverse of the spike period) of nanowire neurons as a function of $I_{in}$ and $I_{bias}$. These tuning currents determine the firing potential of the neuron.}
    \label{fig:3}
\end{figure*}

In spiking neural networks, information must be encoded in the timing of the spikes in the network. In figure \ref{fig:3}, we demonstrate the time-domain response of the neuron circuit with respect to its current biasing conditions. We define the spike period as the time between the voltage spikes in a neuron. Figure \ref{fig:3}a demonstrates a decrease in the spike period of the nanowire neuron as a function of increasing input current. This same behaviour is seen in overbiased relaxation oscillators, where biasing a nanowire above its critical current also results in frequency-tunable oscillations \cite{toomey_superconducting_nodate}. We map the effect of this current-controlled frequency-tunability in figure \ref{fig:3}b. From the color map a firing threshold for the nanowire neuron can be identified from the summation of the input and bias currents. Once this threshold is reached, either an increase in the bias current or an increase in the input current result in an increase in the firing rate. This behavior can be explained  from the critical current dynamics of the nanowires in the neuron circuit.

\begin{figure*}
    \centering
    \includegraphics[width=\textwidth]{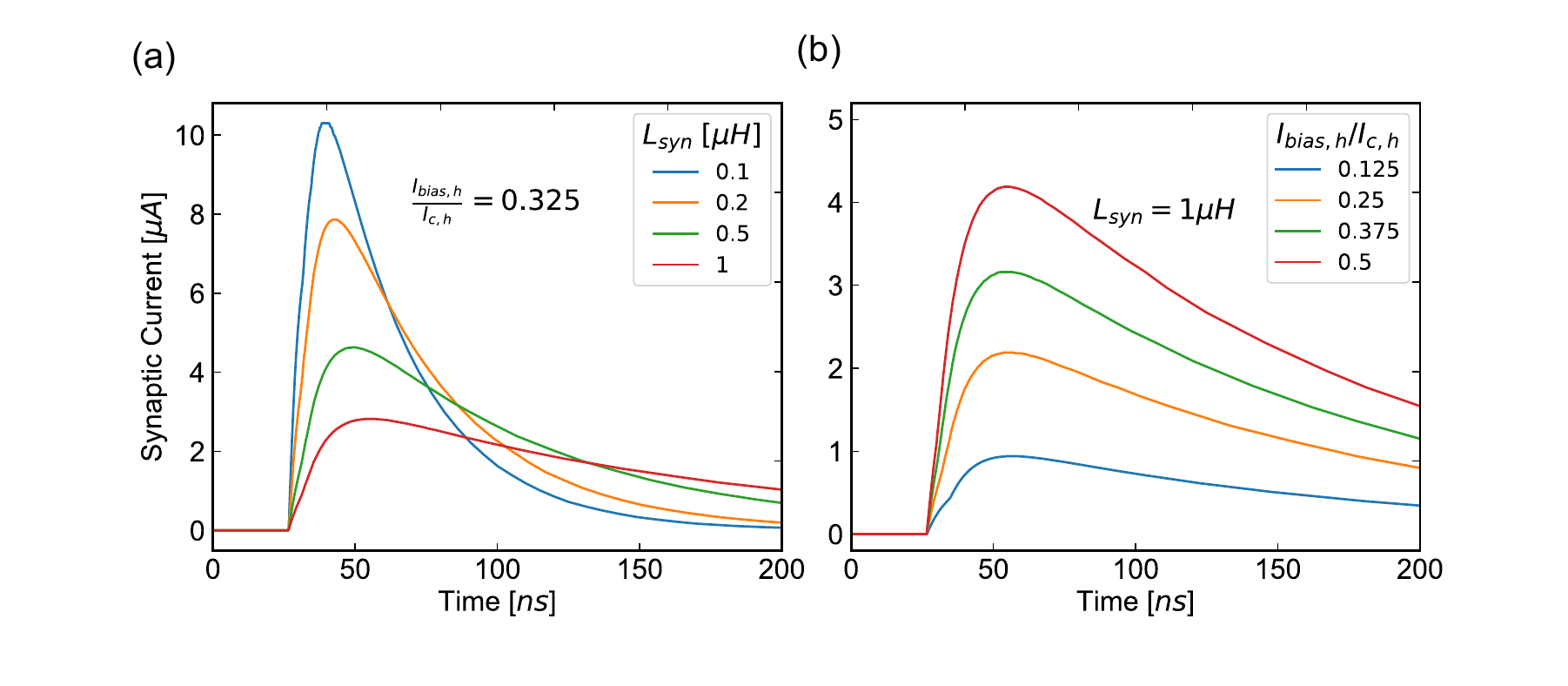}
    \caption{Simulation results of the design space of the hTron synapse and the current-controlled tunability of the hTron synapse. (a) The time-domain response of the synaptic current plotted for various values of $L_{syn}$ (b) The tunability of the synapse strength for various values of $I_{bias,h}$ for $L_{syn} = 1 \; \mu H$. This plot is inverted for negative $I_{bias,h}$.}
    \label{fig:4}
\end{figure*}

Similarly, the synapse circuit presented in figure \ref{fig:1} also demonstrates tunable characteristics. The integration loop of the synapse acts as a leaky integrator circuit. The time-domain response of this circuit can be set during its design, by choosing the ratio between the $L_{syn}/R_{syn}$ time constant and the $L_{nw}/R_2$ time constant, where $L_{nw}$ is the inductance of $nw_2$. The leakiness of the synapse is illustrated in figure \ref{fig:4}a where the current in the synaptic inductor is plotted for different values of synaptic inductance. In addition,  $I_{bias,h}$ controls the amount of current injected into the integration loop, modifying the output of the synapse. Therefore, the strength of the  synaptic connection can be updated externally by tuning $I_{bias,h}$. Figure \ref{fig:4}b shows the output current of the synapse for different values of  $I_{bias,h}$.

\section{Results}
\noindent
To connect our hardware with the computational picture of a neural network, we developed two mappings from hardware to mathematical models. First is a mapping between the physical system described in the previous section and the well-known leaky integrate-and-fire neuronal model \cite{teeter_generalized_2018}, second is the mapping to a recently developed compositional model for spiking neural networks \cite{lynch_basic_2021}.

\subsection{Correspondence}

\subsubsection{Correspondence to Leaky Integrate-and-fire Model}
\noindent
The Leaky integrate-and-fire neural network is one of the most commonly studied network-level models in neuroscience. In a network of $n$ neurons, each is associated with a time-varying potential value. In this model, a neuron’s potential is governed by the following equation:
\begin{equation} 
 \frac{d u_{i}(t)}{d t}=-\lambda\left(u_{i}(t)-u_{0, i}\right)-\sum_{j} \alpha C_{i j} s_{j}(t)+I_{i}(t)
\end{equation}
where $u_i(t)$ represents the potential of the \emph{i}th neuron at time $t$; $u_{0,i}$ is the initial potential with no input; and $I_i(t)$ is the external input to the neuron. The leaky integration dynamic is encapsulated in the leakiness parameter $\lambda$ of the neuron, that is the rate at which the neuron potential decreases. The synaptic strength from a neuron $i$ to a neuron $j$ is represented by matrix coefficient $C_{ij}$ and a transfer parameter $\alpha$. The spiking events of neuron $i$ occur according to the following spike rule $s(t)$:
\begin{equation}
    \begin{aligned}
&s(t)=1, u_{i}(t)>\eta \\
&s(t)=0, \text { otherwise }
\end{aligned}
\end{equation}
where $\eta$ is the threshold potential. We relate the parameters of this model to our superconducting hardware as described below. 

In our hardware, the current in the nanowire $i_{nw}(t)$ of the neuron’s main oscillator corresponds to the potential $u_i(t)$ in the leaky integrate-and-fire model. In the nanowire-based implementation, the spike rule corresponds to the voltage in the nanowire, the spiking events of the nanowire neuron are described by the following spike rule:  
\begin{equation}
   \begin{aligned}
&v_{n w}(t)=i_{n w} R_{h s}, i_{n w}(t)>I_{c} \\
&v_{n w}(t)=0, \text { otherwise.}
\end{aligned} 
\end{equation}

This relatively natural correspondence is what makes the superconducting system particularly elegant for the implementation of spiking neural networks.

Similarly, the initial potential $u_0$ directly corresponds to the initial value of $i_{nw}$. For instance, when the inductance and resistance values between the two branches of the neurons are the same, the bias current in the nanowire of the main oscillator will be $I_{bias}/2$. 

Moreover, the integration behaviour of the model corresponds to the ability of the hTron synapses to integrate the voltage spikes generated by the upstream neurons. The switching of the hTron channel and its subsequent diversion of current into the integration loop of the synapse, can be approximated by a leaky integration circuit with dynamics described by:
\begin{equation}
    \frac{d i_{i n t}}{d t}=\frac{1}{\tau}\left(v_{i n}-i_{i n t}\right) .
\end{equation}

Continuing with the analogy, the leakiness parameter $\lambda$  corresponds to the ratio of the time constant of the relaxation of the nanowire in the neuron to the time constant of the integration loop in the synapse $\tau_{n w} / \tau_{s y n}=\left(L_{n w} / R_{2}\right) /\left(L_{s y n} / R_{s y n _1}\right)$. $\tau_{syn}$ can be set such that it is larger than $\tau_{nw}$ to allow the synapse to retain the spike information from the neurons. 

The transfer parameter $\alpha$ in the model corresponds to the ratio between $I_{bias,h}$ and the current entering the nanowire of the main oscillator. The value of $\alpha$ is dependent on the number of synapses connected to a neuron, the synaptic inductance, and the inductances $L_1$ and $L_2$ of the nanowire neuron. For a neuron with a large number of synapses connected to its input terminal, less current from each synapse enters the neuron.

The matrix coefficients $C_{ij}$ correspond to $I_{bias,h}$ between the \emph{i}th and \emph{j}th neurons as it represents the strength of the connection between two neurons via a synapse. In the hardware, $C_{ij}$ can be externally tuned as discussed in the previous section and illustrated in figure \ref{fig:4}. Coupled together, $C_{ij}$ can be mapped to the current added to the nanowire of the main oscillator.

The $I_i(t)$ terms in the model correspond to the external input current source $I_{in}$ to the \emph{i}th neuron as shown in figure \ref{fig:1}. 

\subsubsection{Correspondence to a Basic Compositional Model}
\noindent
An algorithmic model for spiking neural networks has been recently introduced.\cite{lynch_basic_2021} The model lays out a schema to track a set of neurons (nodes) $V$ with a set of synapses (edges) $E$, in a graph by recording a potential for each neuron, at some discrete time. 
\begin{itemize}
    \item In the model, a neuron can be in one of two states: firing and not firing. For a neuron at node $u$ we have $C_t(u)=1$ when the neuron is firing and $C_t(u)=0$ when not firing. We call this function the \emph{configuration} of a neuron. 
    \item The connection between a neuron $u$ and neuron $v$ via a synapse is encapsulated in a function $w(u,v)$.
    \item At discrete time $t$, every neuron $u$ has a potential $pot_t(u) = [\sum_{(v, u) \in E}C_t(v)w(v,u)] - b(u)$.
    \item The firing rule for the neuron is probabilistic and is described by $p_t(u) = \frac{1}{1 + e^{\frac{-pot_t(u)}{\lambda}}}$. Here $\lambda$ is distinct from the leakiness parameter of the leaky integrate-and-fire model and is instead an arbitrary real \emph{temperature} parameter. 
\end{itemize}

The neuron biasing condition $b(u)$ is akin to the currents $I_{in}$ and $I_{bias}$ into a neuron as in figure \ref{fig:1}. We associate the neuron biasing condition to be the location $(I_{in}, I_{bias})$ as in figure \ref{fig:3}b. 

The configuration of a neuron $C_i(u)$ corresponds to the state of the nanowire of the main oscillator of the neuron. When the nanowire switches, the neuron is firing. Conversely, when the nanowire is in the superconducting state the neuron is not firing. 

The weight of a synapse between two neurons $w(u,v)$ is mapped to $I_{bias,h}$ of the synapse between them. As in the leaky integrate-and-fire model, this weight can be externally tuned as demonstrated in figure \ref{fig:4}. 

The potential of the neuron $pot_t(u)$ can be associated with the current in the nanowire of the main oscillator $i_{nw}(t)$. As shown in figure \ref{fig:1}, this current dependent on the on $I_{bias}$ and $I_{in}$ and is affected by the current coming from the connections of other synapses.

The firing rule $p_t(u)$ can be associated with the switching of the nanowire in the main oscillators. While it is true that the nanowire switches whenever $i_{nw}(t)>I_c$, this is probabilistic in a physical implementation and is dependent on $I_{in}$ and $I_{bias}$. The firing probability as a function of $I_{in}$ and $I_{bias}$ was explored in experiments with nanowire neurons here \cite{toomey_superconducting_2020}.

We summarize the correspondence between the two models presented and the physical parameters in the table I and II. 

\begin{table}
\begin{tabular}{|l|l|}
\hline
Leaky integrate-and-fire model & Physical Model \\
\hline\hline
Initial potential of a neuron $u_{0,i}$ & Initial current in the nanowire, $i_{nw}(0)$ \\ 
\hline
Spike function $s(t)$ &
Voltage spikes at $V_{out}$ \\
\hline
Connection between neurons, $\alpha C_{ij}$ & Bias current in the synapse $I_{bias,h}$ \\
\hline
Potential of a neuron, $u_i(t)$ & Current in the nanowire of the main oscillator $i_{nw}(t)$ \\
\hline
Spike rule $u_i(t)>\eta$ & Nanowire switches when $i_{nw}(t)>I_c$ with noise \\
\hline
Leakiness parameter, $\lambda$ & Time constants of the neuron and synapse $\tau_{nw}/\tau_{syn}$ \\
\hline
\end{tabular}  

\caption{\label{tab:table1} Correspondence between the leaky integrate-and-fire model and the hardware description.}
\end{table}

\begin{table}
\centering
\begin{tabular}{|l|l|}
\hline
Compositional Model  & Physical Model\\
\hline\hline
Bias conditions of a neuron, $b(u)$ & $I_{in}$ and $I_{bias}$ in figure \ref{fig:3} \\ 
\hline
Configuration of a neuron, $C_t(u)$ &
Voltage spikes at $V_{out}$ \\
\hline
Weight of a synapse, $w(u,v)$ & Bias current in the synapse, $I_{bias,h}$ \\
\hline
Potential of a neuron $pot_t(u)$ & Current in the nanowire of the main oscillator $i_{nw}(t)$ \\
\hline
Firing probability, $p_t(u)$ & Nanowire switches when $i_{nw}(t)>I_c$ with noise\\
\hline
\end{tabular}  

\caption{\label{tab:table2} Correspondence between the compositional model and the hardware description.}
\end{table}

\subsection{Model and Translational Tool}
\noindent
From the above descriptions of the leaky-integrate-and-fire model and the basic compositional model for spiking neural networks, we built a tool to directly relate the parameters of the models to the physical implementation of the nanowire neuron and synapse. This tool can help bridge the expertise gap between computer scientists and hardware engineers in designing neuronal circuits as it provides a platform for a common description of a problem.

\subsubsection{Implementation of the Translational Tool}

In the tool, the network consisting of the neurons and synapses is described as a graph. A vector \emph{V} describing the bias conditions to the neurons (the vertices) and a matrix \emph{E} describing the strength of the synapses between them (the edges) is specified. Then, the algorithmic description from the leaky integrate-and-fire model, or the compositional model is chosen. Depending on the choice of model, \emph{V} is treated as $I_i(t)$ or $b(u)$ and \emph{E} is treated as $C_{ij}$ or $w(u,v)$.
Correspondingly, the parameters of the algorithmic model are tuned either at each individual node or across the whole graph.
The tool translates the parameters of each algorithmic model to the low-level hardware description based on the correspondence between the parameters described in each the previous section. 






The tool uses SciPy’s numerical solver IVP \cite{2020SciPy-NMeth} to simulate the underlying system based on a state variable description of the nanowire neuron circuit, as follows: 

\begin{equation}
    \begin{gathered}
\frac{d i_{1}}{d t}=\left(i_{2} R_{1}-i_{1} R_{h s} n_{1}\right) / L_{n w}\left(i_{1}\right) \\
\frac{d i_{2}}{d t}=-\frac{1}{L_{1}+L_{2}}\left(L_{1} \frac{d i_{in}}{d t}+i_{2} R_{1}-i_{4} R_{2}\right)-\frac{d i_{1}}{d t} \\
\frac{d i_{3}}{d t}=\left(i_{4} R_{2}-i_{3} R_{h s} n_{2}\right) / L_{n w}\left(i_{3}\right) \\
\frac{d i_{4}}{d t}=\frac{1}{L_{1}+L_{2}}\left(L_{1} \frac{d i_{in}}{d t}+i_{3} R_{1}-i_{4} R_{2}\right)-\frac{d i_{3}}{d t}
\end{gathered}
\end{equation}

\noindent
in this case, $L_{nw}(i)$ is a nonlinear function accounting for the kinetic inductance of the nanowire following the expression from \cite{berggren_superconducting_2018}. We define the current $i_{in}$ as $I_{i n}+\sum_{k} i_{s y n, k}$ where $i_{s y n, k}$ is the current flowing through $R_{out}$ from each synapse to the neuron and $I_{in}$ is as in figure \ref{fig:1}. The remaining current variables are further defined in the appendix. 

Note that here we make a simplifying assumption about the dynamics of a superconducting nanowire. We specify a state variable $n_i$ for each nanowire to capture whether the nanowire is in the superconducting $(n_i=0)$ or the normal $(n_i=1)$ state. The transition from the superconducting to the normal state is brought about when $i_{nw}(t)>I_c$. The transition form the normal state back to the superconducting state occurs when $i_{nw}(t)<I_r$. 

Similarly, a state-variable description for the synapse is as follows:

\begin{equation}
    \begin{gathered}
\frac{d i_{1}}{d t}=\left(i_{2} R_{s y n, 1}-i_{1} R_{h s} h\right) / L_{n w, h}\left(i_{1}\right) \\
\frac{d i_{2}}{d t}=-\frac{d i_{1}}{d t}-\frac{d i_{3}}{d t} \\
\frac{d i_{3}}{d t}=\left(i_{2} R_{s y n, 1}-i_{4} R_{s y n, 2}\right) / L_{s y n} \\
\frac{d i_{4}}{d t}=\frac{d i_{3}}{d t}-\frac{d i_{5}}{d t} \\
\frac{d i_{5}}{d t}=\left(i_{4} R_{s y n, 2}-i_{5} R_{o u t}\right) / L_{2} .
\end{gathered}
\end{equation}
We use the same simplifying assumption about the dynamics of the channel of the hTron as we use for the nanowires in the neuron. Transitions to the normal state in the channel of the hTron are brought about after its current surpasses $I_{c,h}$ and its return to the superconducting state occurs when its current is below $I_{r,h}$. To couple neuron and the synapse, we force the hTron channel to switch everytime the neuron fires, that is we set $h=1$ whenever $n_2=1$. In an effort to facilitate broad use of this model, we reference the code implementing it here \footnote{https://github.com/qnngroup/neuron}.

As an example, using the tool to relate a leaky integrate-and-fire model to the hardware, the following steps are taken to translate the algorithmic description to the hardware:

\begin{enumerate}
    \item The neurons and the synapses between them are configured as specified by the user in their graph description
    \item By default, the internal parameters of a neuron $\left(L_{n w}, L_{1}, L_{2}, R_{1}, R_{2}\right)$ are set to typical values $\left(10 n H, 20 n H, 20 n H, 5 \Omega, 5 \Omega\right)$ as are the parameters of a synapse \\ $\left(L_{n w, h}, R_{s y n, 1}, R_{s y n, 2}, R_{o u t}\right)$ are set to $(100 n H, 10 \Omega, 10 \Omega, 5 \Omega)$.
    \item $u_0$ and $\eta$ are respectively mapped directly to $I_{bias}$ in the neuron and $I_c$ of the nanowires.
    \item $L_{syn}$ is set such that the leakiness parameter $ \lambda=\left(L_{n w} / R_{2}\right) /\left(L_{s y n} / R_{s y n}\right)$.
    \item For each synapse, $I_{bias,h}$ is set such that the current in $nw_2$ increases by a factor of $C_{ij}$. This ratio is maintained across all synapses. Correspondingly, $I_{c,h}$ is set to be higher than $I_{bias,h}$.
    \item External inputs to the neurons $I_i(t)$ are proportionally mapped to the input currents $I_{i}$ of each neuron.
    \item The network is simulated by solving the IVP's of underlying circuits.
\end{enumerate}

In the following section apply the correspondence of our hardware to two algorithmic examples. We simulate Boolean gates and solve special linear systems with our superconducting-nanowire-based neuromorphic architecture. These choices stem from the ubiquitous nature of Boolean gates and algorithms to solve linear systems in classical computing. 

\subsection{Solving Linear Systems}
\noindent
In a recent paper by Chou et al. \cite{chou_algorithmic_2018}, non-leaky integrate-and-fire neural networks were shown to efficiently solve linear systems. Here, we demonstrate the computational power of SNNs in simulation by implementing their theoretical models using our superconducting nanowire-based architecture. As a proof of concept, we start with solving a simple two-dimensional  linear system. Then, we scale up the problem to a five-dimensional linear system with Laplacian structure.

The motivation for Laplacian linear systems is two-fold:  (1) many practical applications and engineering problems rely on solving large Laplacian linear systems such as diffusion models, graph models and random walks;  (2) some Laplacian linear systems have infinitely many solutions. Chou et al. \cite{chou_algorithmic_2018} predicted that a SNN will converge to the solution with the least vector magnitude. We use the approach taken by Chou et al. \cite{chou_algorithmic_2018} to solve Laplacian linear systems, namely a 2$\times$2 and a 5$\times$5 system.

\begin{figure}
    \centering
    \includegraphics[width=\textwidth]{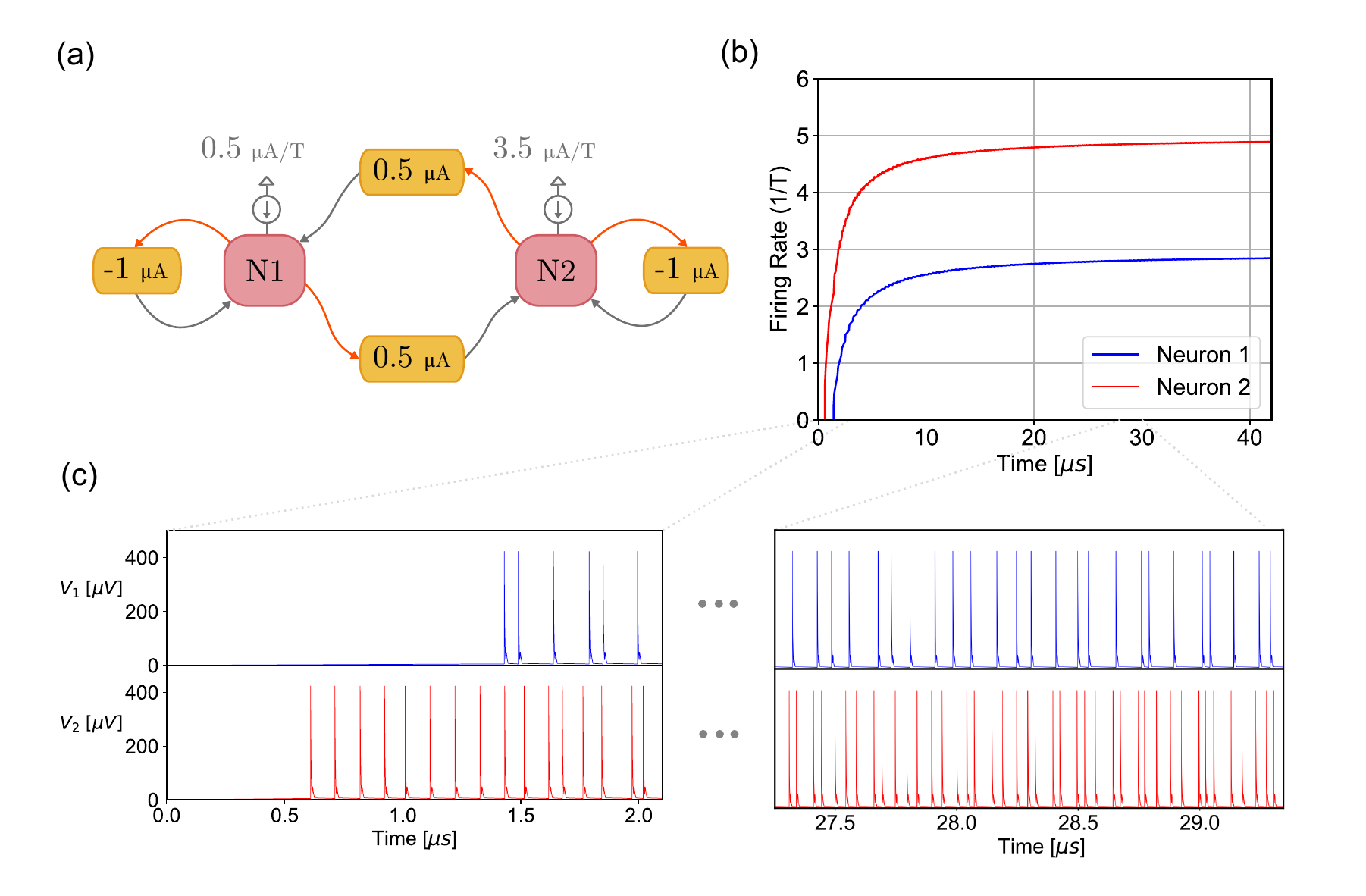}
    \caption{Implementation of an SNN to solve a linear system using simulated superconducting hardware. (a) Graph representation of a network with two neurons (N1, N2) with synapses. The weight of each synapse is inscribed in the synapse itself  (b) Plot of the calculated firing rate of each neuron as the system evolves. (c) Voltage spike waveforms for each neuron at its output voltage node. For this simulation the timescale $T$ of a neuron spike is approximately $37 ns$. Note that orange arrows represent connections via the thermal domain and grey arrows represent connections via the electrical domain}
    \label{fig:5}
\end{figure}

To translate a linear system of the form $Ax=b$ to a SNN, we map $C_{ij}$ to be the elements of the matrix $A^TA$ and $I(t)$ to be the vector $A^Tb$ to ensure that the matrix $C$ is positive semidefinite (PSD). The number of neurons in the SNN corresponds to the dimension of $A$.

For the first example, we attempt to solve the following $Ax=b$ linear system which is already PSD:

\begin{equation}
    \left[\begin{array}{cc}
1 & -0.5 \\
-0.5 & 1
\end{array}\right] x=\left[\begin{array}{l}
0.5 \\
3.5
\end{array}\right]
\end{equation}

We illustrate the network for solving this system and use the tool to handle parameter mapping as in figure \ref{fig:5}. The connectivity matrix $C_{ij}$ from the leaky integrate-and-fire model corresponds to the matrix $(-1)\times A$ in the problem. We can then map the elements of matrix $A$ from the problem to the weights of the synapses as shown in figure \ref{fig:5}a. Similarly, the row elements of vector $b$ are mapped to the ramp rates of currents at $I_{in}$ for each of the neurons relative to the timescale $T$ of the neuron. To apply the leaky integrate-and-fire model, we set the timescale of integration $\lambda=0.02$ to ensure that the current in the synapse decays much slower relative to the decay of current after a neuron spike. Using the tool, we chose $\alpha=0.67$ and $u_0=0.95\eta$.

The evolution of the system is illustrated in figure \ref{fig:5}c. Initially, neither neuron is firing. As time progresses, the input current to each neuron increases at different rates. Since the external bias for neuron 2 is greater, its potential will increase faster and it will fire earlier. When neuron 2 fires, both of its outgoing synapses are activated. Neuron 2 excites neuron 1 but also inhibits itself. After some time, neuron 1 will fire as a result of the excitation from neuron 2. When neuron 1 fires, it will excite neuron 2 but inhibit itself. As can be seen in figure \ref{fig:5}b, two distinct firing rates emerge from the system. Specifically, we reference the approach of Chou et al. \cite{chou_algorithmic_2018} to define a firing rate as:

\begin{equation}
    N(t)/t
\end{equation}
\noindent
where $N(t)$ is the cumulative number of spikes at time $t$. As can be seen in figure \ref{fig:5}b, we find that the firing rate of each neuron converges to the rows of the solution vector of the linear system $x = [3\;5]^T$.

To illustrate the generality of the method, we extend the approach to solving a more complex linear system. We apply the same approach to solving a cycle graph represented by the following $Ax=b$ linear system which is again already PSD:
\begin{equation}
    \left[\begin{array}{ccccc}
1 & -0.5 & 0 & 0 & -0.5 \\
-0.5 & 1 & -0.5 & 0 & 0 \\
0 & -0.5 & 1 & -0.5 & 0 \\
0 & 0 & -0.5 & 1 & -0.5 \\
-0.5 & 0 & 0 & -0.5 & 1
\end{array}\right] x =\left[\begin{array}{c}
-2.5 \\
0 \\
0 \\
0 \\
-2.5
\end{array}\right]
\end{equation}

\begin{figure}
    \centering
    \includegraphics[width=\textwidth]{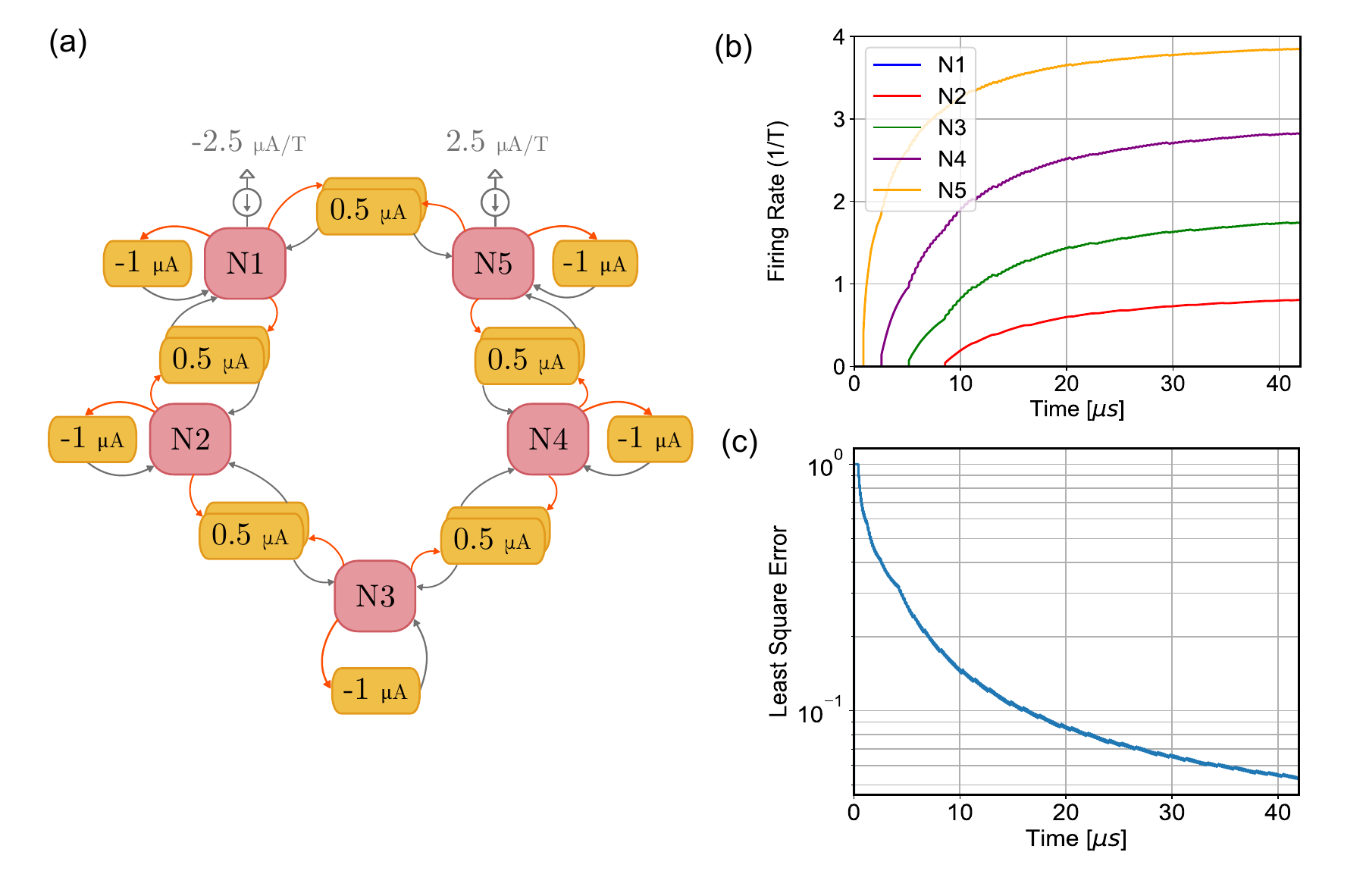}
    \caption{Implementation of an SNN to solve a cycle graph using simulated superconducting hardware. (a) Graph representation of a network with two neurons (N1-N5) with synapses. The weight of each synapse is inscribed in the synapse itself (b) Plot of the firing rate of each neuron as the system simulation evolves. (c) Calculated least square error of the solution as the system evolves. For this simulation, the timescale of a neuron spike is approximately $31 \text{ns}$. Note that orange arrows represent connections via the thermal domain and grey arrows represent connections via the electrical domain}
    \label{fig:6}
\end{figure}

We can map again the elements of matrix $A$ to the connectivity matrix $C_{ij}$ and the elements of vector $b$ to the input currents of the neurons. We use the tool described in the previous section to simulate the system and illustrate the results in figure \ref{fig:6}. 

These results can be understood from an energy minimization perspective. When neuron 5 fires, it will excite neuron 4 after a long period of time. When neuron 4 begins firing continuously, neuron 3 will be excited. This same effect will propagate to neuron 2 after a period of time. Neuron 1 will not fire due to the fact that the input current to the neuron is negative. This is illustrated in figure \ref{fig:6}b where the firing rates of the different neurons have different activation times. It must then be noted that this linear system $Ax=b$ defined by the cycle graph has multiple solutions and the firing rate of our SNN approaches the solution with the least L1 norm as predicted in \cite{chou_algorithmic_2018}: $x = \left[
0 \; 
1 \;
2 \;
3 \;
4
\right]^T$

To assess the evolution of the network and determine error in the firing rate, we defined the least square error as follows:
\begin{equation}
    \operatorname{err}=\|A x-b\| /\|b\|
\end{equation}

Where the notation $||b||$ signifies the magnitude of the vector. We take $x$ to be the instantaneous firing rate vector. We plot the evolution of the least square error in figure \ref{fig:6}c.

\subsection{Boolean Gates}
\noindent
We also implement several Boolean gate network examples found in Lynch and Musco’s paper \cite{lynch_basic_2021}. The networks were created using their algorithmic model and then translated into our neuromorphic hardware. Boolean gates are relevant algorithmic examples to convert into neuromorphic computing hardware given their high importance in classical computing and their use in neural networks. Thus, neuromorphic versions of a universal set of Boolean gates could enable computation with both classical and neuromorphic paradigms.

We demonstrate a 3-input AND gate network in figure 7. In the following paragraph, we describe the operation of this network using the compositional framework from Lynch and Musco’s paper \cite{lynch_basic_2021}. We understand the operation of the network using the compositional model. The weight of the synapses are taken to be $L$. When all three of the input neurons fire, the potential of the neuron is $-b+3L$ and the probability of the output neuron firing is $(1+exp(b-3L))^{-1}$. When only two input neurons fire, the probability of the output neuron firing is $(1+exp(b-2L))^{-1}$. If we take $L=2 \ln(\frac{1-\delta}{\delta})$ and $b=\frac{5}{2}L$ we can see that the probability of output neuron firing when all three input neurons fire is $1-\delta$ for $\delta$ being an arbitrary small parameter. In practice, the synapse and neuron biasing conditions determine $\delta$, allowing $\delta$ to be set arbitrarily close to 0. In a physical implementation of the network, the probability of the output neuron firing can be attributed to current noise which causes fluctuations in the current of the nanowires. If a nanowire is biased very closely to $I_c$ then there is a probability it may switch.

\begin{figure}
    \centering
    \includegraphics[width=\textwidth]{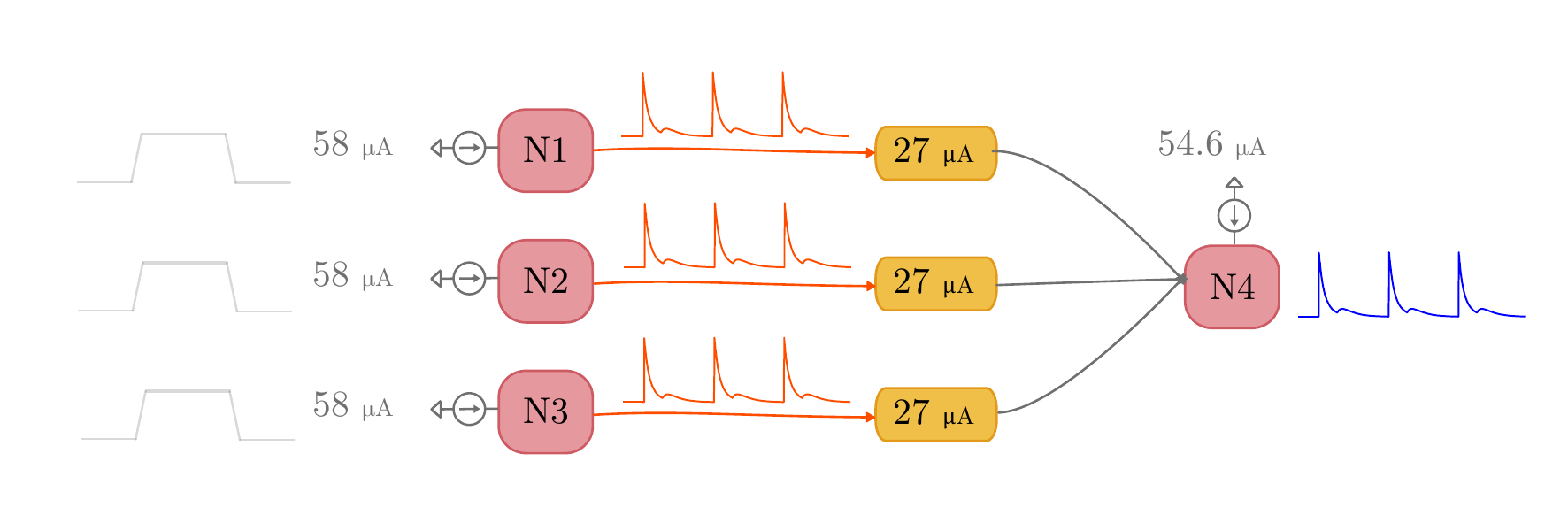}
    \caption{3-input AND gate. The synapse bias current for each connection is $27\mu A$, the input neuron bias current is $58 \mu A$, and the corresponding bias for the output neuron is $54.6\mu A$ for 3 inputs. The critical current for nanowires in the neurons is $I_c=30 \mu A$. The current from an input neuron must be greater than $3.72 \mu A$ for it to fire. As more inputs are added the output neuron bias would have to be lowered accordingly.}
    \label{fig:7}
\end{figure}

\begin{figure}
    \centering
    \includegraphics[width=\textwidth]{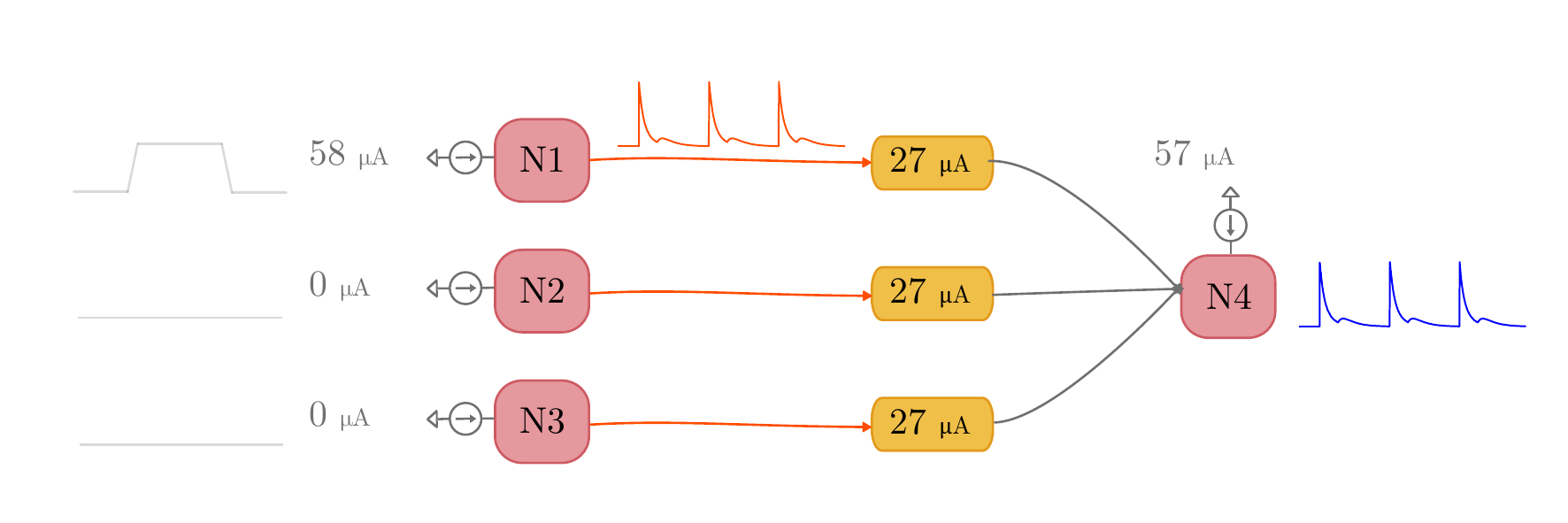}
    \caption{3-input OR gate. The synapse bias current for each connection is $27\mu A$, the input neuron bias current is $58 \mu A$, and the corresponding bias for the output neuron is $57\mu A$ for 3 inputs. The critical current for nanowires in the neurons is $I_c=30 \mu A$. This output bias would not have to be lowered upon adding inputs because it only needs to fire if it receives enough input current from any one of the synapse connections.}
    \label{fig:8}
\end{figure}

In addition, we demonstrate a 3-input OR gate in figure \ref{fig:8}. The input currents and the synapse bias currents are similar to the AND gate. Through an analogous formulation as in the AND gate, we can set the biases in the network such that the output neuron fires with probability $1 - \delta$ when one of the input neurons fires, it fires with probability $\delta$ if none of the input neurons fire. Again, $\delta$ can be made arbitrarily close to 0 in this model. We can understand this as a threshold problem allowing for the robust implementation of Boolean gates. 

\section{Discussion}
\noindent

Solving a physical system by using another physical system with similar dynamics as its model is a promising approach to computing.  Here we discuss the advantages and failings of the approach and provide insight on how such a system could be realized. 

\subsection{Solving Linear Systems with Nanowire Neurons}
\noindent
The approach by Chou et al. for solving linear systems implemented in this paper is numerically robust. The solution of the $Ax=b$ system is not physically tied to an experimental observable, such as voltage or current. As in \cite{chou_algorithmic_2018}, the numerical accuracy is a function of the total evolution time of the system. Accuracy can thus be optimized by increasing the time as well as by decreasing the time scale of the components of the circuits. Superconducting nanowires offer rise times on the order of a few ps and relaxation times as short as 2-5 ns from previous measurements \cite{toomey_superconducting_nodate}. Thus, least-square errors below $10^{-3}$ might be achieved within tens of microseconds for large networks. The values of the resistors and inductors chosen to set the $L/R$ time constants with the tool do not constraint the time scale of the SNN . Therefore, the solution can be obtained independently of the timing parameters set in the system. 
The implementation of this approach with superconducting hardware remains to be demonstrated experimentally. In neurons with high fan-in, that is with many incoming connections from synapses, the resistive network connecting a set of synapses to the input terminal of the neuron can result in significant power consumption. Higher current and resistances are needed to account for leakage current in such a network. Some previous approaches \cite{castellani_design_nodate,primavera_active_2021} have suggested the use of fan-in trees to mitigate this problem which could be beneficial in our architecture. However, using either resistive networks or tree structure would still pose problems with either power or area scaling. An improved architecture for fan-in is needed to break this scaling concern.

The advantage of our architecture is the decoupling of the neuron output to the synapse input via the hTron. Since no electrical connection is needed between the output node of a neuron to a synapse, there is no need for impedance matching. Hence, large fan-out can be achieved by patterning $R_2$ to allow for heat dissipation at multiple locations where the hTron channels of multiple synapses could be located. However, this of course comes at a cost of higher power needed. An estimation of the power consumption of an implementation of the nanowire neuron can be found in previous work \cite{toomey_design_2019}.
 
 Due to non-idealities in the fabrication process for superconducting electronics, there can be difficulty in achieving small variance in the resistances, inductances, and critical currents of circuit components. This can lead to a spread in the distribution of the spike rise and relaxation times. However, variance across neurons is masked by the definition of the firing rate. After the network evolves for an appreciable time, the time between the spikes need not matter more than the total number of spikes. The firing rate is thus a robust quantity. Even with non-idealities in the fabrication process, the architecture proposed in this work would not be significantly different from biological neurons and synapses, which intrinsically have variability. Here, we make the assumption that the design of the hTron can be optimized such that the synapses are still activated despite variations in the strength of the voltage spike at the neuron.

While some simple checks are implemented to check parameter ranges that are practically realizable by fabrication, additional experimental verification may still be needed. For instance, the range of synaptic weights in the synapses that can be designed via the inductance of $L_{syn}$ is limited by the kinetic inductance of the material used. We have chosen NbN with kinetic inductance $~ 33\; pH/sq$ for our tool owing to previous reports of implementations and measurements of nanowire neurons and synapses \cite{toomey_design_2019, castellani_design_nodate}. To alleviate this problem, higher-kinetic-inductance materials such as WSi with kinetic inductance of $~260\; pH/sq$ \cite{mccaughan_kinetic-inductance-based_2018} can be chosen instead. Achieving lower resistances is limited by the presence contact resistance and the inherent variability in fabrication processes. Similarly, the range of synaptic weights is also limited by the design of the hTron. The upper bound for the value of $I_{bias,h}$ stems from $I_{c,h}$. Superconducting nanowires can be designed to have critical currents above 1 mA. However, challenges could arise with the size of components on chip. Larger critical currents typically are obtained from an increase in cross-sectional area of the superconducting trace. As a result, there is a large area cost for synaptic inductors with large kinetic inductances and high critical currents. This constraint may limit the values of $C_{ij}$ that could be realized. 

Another consideration in the design of the circuit network is the readout of the firing rate. For a fully integrated system as in \cite{davies_loihi_2018-1}, we envision a multi-layered system with readout and control circuitry  connected by vias. For the firing rate readout circuit, superconducting counter circuits could be implemented based on nTron devices \cite{mccaughan_superconducting-nanowire_2014}. Similarly, a hybrid superconductor-transistor approach could allow for readout via classical digital logic \cite{xie_nbn-gated_2021}. In cases where the spike strength or the weights of the synapses are not known, the linearity of an $Ax=b$ system could be exploited. Instead of using the individual firing rates of the neurons as the components of the solution vector $x$, the ratio of the firing rates of the neurons could be used instead. Superconducting single flux quantum (SFQ) logic architectures would then be available to realize more complex readout circuits \cite{pasandi_sfqmap_2019} while still offering the advantages of superconducting electronics. 

\subsection{Modelling}
\noindent
We have taken a simpler approach to modelling superconducting elements to accommodate the possibility of increased scaling. The tool ignores the microscopic electrodynamics of the system and the rigorous electrothermal physics that describes device operation. Instead, the state-variable description of the switching of the nanowires in the neuron and the hTron channel in the synapses is a simplification of the phenomenological models of \cite{berggren_superconducting_2018, castellani_design_nodate} which is convenient for algorithm designers and avoids non-linearities in the model \cite{baghdadi_multilayered_2020}. For larger nanowires, there may be non-trivial dynamics that would deviate from the lumped element model and discrepancies that could arise from the temperature dependence of the physical parameters.

In the first example presented in the previous section, the tool enabled the translation of algorithmic parameters from the leaky integrate-and-fire model $\left(\lambda, \alpha, C_{i j}, I_{i}\right)$ into specifications for the hardware $\left(R_{1,2}, L_{1,2}, R_{s y n, 1,2}, L_{s y n}, R_{o u t}, \text { etc. }\right)$ with ease. While the approach presented in the previous section of choosing the parameters is not necessarily unique, it is possible for experts outside of superconducting electronics to understand and apply. Hence, there is no expertise in superconducting electronics required to explore further applications. Our superconducting hardware and its associated tool is versatile because it can be associated with many computational models---two are shown in this paper. Superconducting nanowires have also been applied in image recognition, Winner-Takes-All algorithms, stochastic behaviour \cite{toomey_superconducting_2020}; and more conventional electronics \cite{zhao_compact_2018}. By the same token, the tool is not dependent on circuit modelling software such as LTSPICE and uses the more common language of python rather than a higher-level professional language like Verilog A. As a result, the functionality of the tool presented in this work can be similarly extended to other superconducting systems based on Josephson junctions \cite{goteti_superconducting_2021,segall_phase-flip_2014,shainline_superconducting_2019}, and quantum phase-slip junctions \cite{cheng_spiking_2018,cheng_toward_2021} albeit with increased complexity for the component models. Optimizing circuit layouts for power or area given a set of algorithmic constraints would also be an area of extension for the tool. These ideas constitute a useful continuation of this work.

\section{Conclusion}
\noindent
We presented the fundamental components for the hardware implementation of a neural network based on superconducting nanowires. We translated the hardware architecture to its algorithmic description enabling a straightforward understanding of the algorithmic correspondence of physical parameters. This understanding elucidates how more complicated networks of arbitrary scale can be built based on robust theoretical models. In addition, the work incites the future implementation of new models for biological neurons and synapses to replicate more complex bio-realistic behaviour. The description of a leaky integrate-and-fire model in terms of physical parameters enables the exploration of the design and fabrication of circuit layouts corresponding to linear system solvers.

Most importantly, the encapsulation of this work in a python-based tool is key in filling the gap between algorithmic designers and hardware designers. It is a point of commonality for the expertise within both of these fields. It can thus enable, in the future, concrete and fast approaches to solving neuromorphic problems using a superconducting nanowire-based neuromorphic architecture. The direct translation of superconducting neuromorphic architectures into algorithmic formulations of a problem is facilitated with this tool. As a result, it is now easier for hardware designers to condense an abstract algorithmic problem into a specific hardware platform without increasing the complexity of circuits or compromising energy efficiency as is typical of CMOS circuits. Thus, the issue is no longer a question of expertise. 

\begin{acknowledgments}
We wish to acknowledge the support of and thoughtful discussions with collaborators. 
\end{acknowledgments}
\section{Data }
\section{References}
\bibliography{main}

\begin{thebibliography}{38}%
\makeatletter
\providecommand \@ifxundefined [1]{%
 \@ifx{#1\undefined}
}%
\providecommand \@ifnum [1]{%
 \ifnum #1\expandafter \@firstoftwo
 \else \expandafter \@secondoftwo
 \fi
}%
\providecommand \@ifx [1]{%
 \ifx #1\expandafter \@firstoftwo
 \else \expandafter \@secondoftwo
 \fi
}%
\providecommand \natexlab [1]{#1}%
\providecommand \enquote  [1]{``#1''}%
\providecommand \bibnamefont  [1]{#1}%
\providecommand \bibfnamefont [1]{#1}%
\providecommand \citenamefont [1]{#1}%
\providecommand \href@noop [0]{\@secondoftwo}%
\providecommand \href [0]{\begingroup \@sanitize@url \@href}%
\providecommand \@href[1]{\@@startlink{#1}\@@href}%
\providecommand \@@href[1]{\endgroup#1\@@endlink}%
\providecommand \@sanitize@url [0]{\catcode `\\12\catcode `\$12\catcode
  `\&12\catcode `\#12\catcode `\^12\catcode `\_12\catcode `\%12\relax}%
\providecommand \@@startlink[1]{}%
\providecommand \@@endlink[0]{}%
\providecommand \url  [0]{\begingroup\@sanitize@url \@url }%
\providecommand \@url [1]{\endgroup\@href {#1}{\urlprefix }}%
\providecommand \urlprefix  [0]{URL }%
\providecommand \Eprint [0]{\href }%
\providecommand \doibase [0]{https://doi.org/}%
\providecommand \selectlanguage [0]{\@gobble}%
\providecommand \bibinfo  [0]{\@secondoftwo}%
\providecommand \bibfield  [0]{\@secondoftwo}%
\providecommand \translation [1]{[#1]}%
\providecommand \BibitemOpen [0]{}%
\providecommand \bibitemStop [0]{}%
\providecommand \bibitemNoStop [0]{.\EOS\space}%
\providecommand \EOS [0]{\spacefactor3000\relax}%
\providecommand \BibitemShut  [1]{\csname bibitem#1\endcsname}%
\let\auto@bib@innerbib\@empty
\bibitem [{\citenamefont {Marković}\ \emph {et~al.}(2020)\citenamefont
  {Marković}, \citenamefont {Mizrahi}, \citenamefont {Querlioz},\ and\
  \citenamefont {Grollier}}]{markovic_physics_2020}%
  \BibitemOpen
  \bibfield  {author} {\bibinfo {author} {\bibfnamefont {D.}~\bibnamefont
  {Marković}}, \bibinfo {author} {\bibfnamefont {A.}~\bibnamefont {Mizrahi}},
  \bibinfo {author} {\bibfnamefont {D.}~\bibnamefont {Querlioz}},\ and\
  \bibinfo {author} {\bibfnamefont {J.}~\bibnamefont {Grollier}},\ }\bibfield
  {title} {{\selectlanguage {english}\bibinfo {title} {Physics for neuromorphic
  computing}},\ }\href {https://doi.org/10.1038/s42254-020-0208-2} {\bibfield
  {journal} {\bibinfo  {journal} {Nature Reviews Physics}\ }\textbf {\bibinfo
  {volume} {2}},\ \bibinfo {pages} {499} (\bibinfo {year} {2020})},\ \bibinfo
  {note} {bandiera\_abtest: a Cg\_type: Nature Research Journals Number: 9
  Primary\_atype: Reviews Publisher: Nature Publishing Group Subject\_term:
  Electronics, photonics and device physics;Nanoscale devices
  Subject\_term\_id:
  electronics-photonics-and-device-physics;nanoscale-devices}\BibitemShut
  {NoStop}%
\bibitem [{\citenamefont {Berggren}\ \emph {et~al.}(2020)\citenamefont
  {Berggren}, \citenamefont {Xia}, \citenamefont {Likharev}, \citenamefont
  {Strukov}, \citenamefont {Jiang}, \citenamefont {Mikolajick}, \citenamefont
  {Querlioz}, \citenamefont {Salinga}, \citenamefont {Erickson}, \citenamefont
  {Pi}, \citenamefont {Xiong}, \citenamefont {Lin}, \citenamefont {Li},
  \citenamefont {Chen}, \citenamefont {Xiong}, \citenamefont {Hoskins},
  \citenamefont {Daniels}, \citenamefont {Madhavan}, \citenamefont {Liddle},
  \citenamefont {McClelland}, \citenamefont {Yang}, \citenamefont {Rupp},
  \citenamefont {Nonnenmann}, \citenamefont {Cheng}, \citenamefont {Gong},
  \citenamefont {Lastras-Montaño}, \citenamefont {Talin}, \citenamefont
  {Salleo}, \citenamefont {Shastri}, \citenamefont {Lima}, \citenamefont
  {Prucnal}, \citenamefont {Tait}, \citenamefont {Shen}, \citenamefont {Meng},
  \citenamefont {Roques-Carmes}, \citenamefont {Cheng}, \citenamefont
  {Bhaskaran}, \citenamefont {Jariwala}, \citenamefont {Wang}, \citenamefont
  {Shainline}, \citenamefont {Segall}, \citenamefont {Yang}, \citenamefont
  {Roy}, \citenamefont {Datta},\ and\ \citenamefont
  {Raychowdhury}}]{berggren_roadmap_2020}%
  \BibitemOpen
  \bibfield  {author} {\bibinfo {author} {\bibfnamefont {K.}~\bibnamefont
  {Berggren}}, \bibinfo {author} {\bibfnamefont {Q.}~\bibnamefont {Xia}},
  \bibinfo {author} {\bibfnamefont {K.~K.}\ \bibnamefont {Likharev}}, \bibinfo
  {author} {\bibfnamefont {D.~B.}\ \bibnamefont {Strukov}}, \bibinfo {author}
  {\bibfnamefont {H.}~\bibnamefont {Jiang}}, \bibinfo {author} {\bibfnamefont
  {T.}~\bibnamefont {Mikolajick}}, \bibinfo {author} {\bibfnamefont
  {D.}~\bibnamefont {Querlioz}}, \bibinfo {author} {\bibfnamefont
  {M.}~\bibnamefont {Salinga}}, \bibinfo {author} {\bibfnamefont {J.~R.}\
  \bibnamefont {Erickson}}, \bibinfo {author} {\bibfnamefont {S.}~\bibnamefont
  {Pi}}, \bibinfo {author} {\bibfnamefont {F.}~\bibnamefont {Xiong}}, \bibinfo
  {author} {\bibfnamefont {P.}~\bibnamefont {Lin}}, \bibinfo {author}
  {\bibfnamefont {C.}~\bibnamefont {Li}}, \bibinfo {author} {\bibfnamefont
  {Y.}~\bibnamefont {Chen}}, \bibinfo {author} {\bibfnamefont {S.}~\bibnamefont
  {Xiong}}, \bibinfo {author} {\bibfnamefont {B.~D.}\ \bibnamefont {Hoskins}},
  \bibinfo {author} {\bibfnamefont {M.~W.}\ \bibnamefont {Daniels}}, \bibinfo
  {author} {\bibfnamefont {A.}~\bibnamefont {Madhavan}}, \bibinfo {author}
  {\bibfnamefont {J.~A.}\ \bibnamefont {Liddle}}, \bibinfo {author}
  {\bibfnamefont {J.~J.}\ \bibnamefont {McClelland}}, \bibinfo {author}
  {\bibfnamefont {Y.}~\bibnamefont {Yang}}, \bibinfo {author} {\bibfnamefont
  {J.}~\bibnamefont {Rupp}}, \bibinfo {author} {\bibfnamefont {S.~S.}\
  \bibnamefont {Nonnenmann}}, \bibinfo {author} {\bibfnamefont {K.-T.}\
  \bibnamefont {Cheng}}, \bibinfo {author} {\bibfnamefont {N.}~\bibnamefont
  {Gong}}, \bibinfo {author} {\bibfnamefont {M.~A.}\ \bibnamefont
  {Lastras-Montaño}}, \bibinfo {author} {\bibfnamefont {A.~A.}\ \bibnamefont
  {Talin}}, \bibinfo {author} {\bibfnamefont {A.}~\bibnamefont {Salleo}},
  \bibinfo {author} {\bibfnamefont {B.~J.}\ \bibnamefont {Shastri}}, \bibinfo
  {author} {\bibfnamefont {T.~F.~d.}\ \bibnamefont {Lima}}, \bibinfo {author}
  {\bibfnamefont {P.}~\bibnamefont {Prucnal}}, \bibinfo {author} {\bibfnamefont
  {A.~N.}\ \bibnamefont {Tait}}, \bibinfo {author} {\bibfnamefont
  {Y.}~\bibnamefont {Shen}}, \bibinfo {author} {\bibfnamefont {H.}~\bibnamefont
  {Meng}}, \bibinfo {author} {\bibfnamefont {C.}~\bibnamefont {Roques-Carmes}},
  \bibinfo {author} {\bibfnamefont {Z.}~\bibnamefont {Cheng}}, \bibinfo
  {author} {\bibfnamefont {H.}~\bibnamefont {Bhaskaran}}, \bibinfo {author}
  {\bibfnamefont {D.}~\bibnamefont {Jariwala}}, \bibinfo {author}
  {\bibfnamefont {H.}~\bibnamefont {Wang}}, \bibinfo {author} {\bibfnamefont
  {J.~M.}\ \bibnamefont {Shainline}}, \bibinfo {author} {\bibfnamefont
  {K.}~\bibnamefont {Segall}}, \bibinfo {author} {\bibfnamefont {J.~J.}\
  \bibnamefont {Yang}}, \bibinfo {author} {\bibfnamefont {K.}~\bibnamefont
  {Roy}}, \bibinfo {author} {\bibfnamefont {S.}~\bibnamefont {Datta}},\ and\
  \bibinfo {author} {\bibfnamefont {A.}~\bibnamefont {Raychowdhury}},\
  }\bibfield  {title} {{\selectlanguage {english}\bibinfo {title} {Roadmap on
  emerging hardware and technology for machine learning}},\ }\href
  {https://doi.org/10.1088/1361-6528/aba70f} {\bibfield  {journal} {\bibinfo
  {journal} {Nanotechnology}\ }\textbf {\bibinfo {volume} {32}},\ \bibinfo
  {pages} {012002} (\bibinfo {year} {2020})},\ \bibinfo {note} {publisher: IOP
  Publishing}\BibitemShut {NoStop}%
\bibitem [{\citenamefont {Lynch}\ and\ \citenamefont
  {Musco}(2021)}]{lynch_basic_2021}%
  \BibitemOpen
  \bibfield  {author} {\bibinfo {author} {\bibfnamefont {N.}~\bibnamefont
  {Lynch}}\ and\ \bibinfo {author} {\bibfnamefont {C.}~\bibnamefont {Musco}},\
  }\bibfield  {title} {\bibinfo {title} {A {Basic} {Compositional} {Model} for
  {Spiking} {Neural} {Networks}},\ }\href {http://arxiv.org/abs/1808.03884}
  {\bibfield  {journal} {\bibinfo  {journal} {arXiv:1808.03884 [cs]}\ }
  (\bibinfo {year} {2021})},\ \bibinfo {note} {arXiv: 1808.03884}\BibitemShut
  {NoStop}%
\bibitem [{\citenamefont {Davies}\ \emph {et~al.}(2018)\citenamefont {Davies},
  \citenamefont {Srinivasa}, \citenamefont {Lin}, \citenamefont {Chinya},
  \citenamefont {Cao}, \citenamefont {Choday}, \citenamefont {Dimou},
  \citenamefont {Joshi}, \citenamefont {Imam}, \citenamefont {Jain},
  \citenamefont {Liao}, \citenamefont {Lin}, \citenamefont {Lines},
  \citenamefont {Liu}, \citenamefont {Mathaikutty}, \citenamefont {McCoy},
  \citenamefont {Paul}, \citenamefont {Tse}, \citenamefont {Venkataramanan},
  \citenamefont {Weng}, \citenamefont {Wild}, \citenamefont {Yang},\ and\
  \citenamefont {Wang}}]{davies_loihi_2018-1}%
  \BibitemOpen
  \bibfield  {author} {\bibinfo {author} {\bibfnamefont {M.}~\bibnamefont
  {Davies}}, \bibinfo {author} {\bibfnamefont {N.}~\bibnamefont {Srinivasa}},
  \bibinfo {author} {\bibfnamefont {T.-H.}\ \bibnamefont {Lin}}, \bibinfo
  {author} {\bibfnamefont {G.}~\bibnamefont {Chinya}}, \bibinfo {author}
  {\bibfnamefont {Y.}~\bibnamefont {Cao}}, \bibinfo {author} {\bibfnamefont
  {S.~H.}\ \bibnamefont {Choday}}, \bibinfo {author} {\bibfnamefont
  {G.}~\bibnamefont {Dimou}}, \bibinfo {author} {\bibfnamefont
  {P.}~\bibnamefont {Joshi}}, \bibinfo {author} {\bibfnamefont
  {N.}~\bibnamefont {Imam}}, \bibinfo {author} {\bibfnamefont {S.}~\bibnamefont
  {Jain}}, \bibinfo {author} {\bibfnamefont {Y.}~\bibnamefont {Liao}}, \bibinfo
  {author} {\bibfnamefont {C.-K.}\ \bibnamefont {Lin}}, \bibinfo {author}
  {\bibfnamefont {A.}~\bibnamefont {Lines}}, \bibinfo {author} {\bibfnamefont
  {R.}~\bibnamefont {Liu}}, \bibinfo {author} {\bibfnamefont {D.}~\bibnamefont
  {Mathaikutty}}, \bibinfo {author} {\bibfnamefont {S.}~\bibnamefont {McCoy}},
  \bibinfo {author} {\bibfnamefont {A.}~\bibnamefont {Paul}}, \bibinfo {author}
  {\bibfnamefont {J.}~\bibnamefont {Tse}}, \bibinfo {author} {\bibfnamefont
  {G.}~\bibnamefont {Venkataramanan}}, \bibinfo {author} {\bibfnamefont
  {Y.-H.}\ \bibnamefont {Weng}}, \bibinfo {author} {\bibfnamefont
  {A.}~\bibnamefont {Wild}}, \bibinfo {author} {\bibfnamefont {Y.}~\bibnamefont
  {Yang}},\ and\ \bibinfo {author} {\bibfnamefont {H.}~\bibnamefont {Wang}},\
  }\bibfield  {title} {\bibinfo {title} {Loihi: {A} {Neuromorphic} {Manycore}
  {Processor} with {On}-{Chip} {Learning}},\ }\href
  {https://doi.org/10.1109/MM.2018.112130359} {\bibfield  {journal} {\bibinfo
  {journal} {IEEE Micro}\ }\textbf {\bibinfo {volume} {38}},\ \bibinfo {pages}
  {82} (\bibinfo {year} {2018})},\ \bibinfo {note} {conference Name: IEEE
  Micro}\BibitemShut {NoStop}%
\bibitem [{\citenamefont {Davies}(2021)}]{davies_lessons_2021}%
  \BibitemOpen
  \bibfield  {author} {\bibinfo {author} {\bibfnamefont {M.}~\bibnamefont
  {Davies}},\ }\bibfield  {title} {\bibinfo {title} {Lessons from {Loihi}:
  {Progress} in {Neuromorphic} {Computing}},\ }in\ \href
  {https://doi.org/10.23919/VLSICircuits52068.2021.9492385} {\emph {\bibinfo
  {booktitle} {2021 {Symposium} on {VLSI} {Circuits}}}}\ (\bibinfo {year}
  {2021})\ pp.\ \bibinfo {pages} {1--2},\ \bibinfo {note} {iSSN:
  2158-5636}\BibitemShut {NoStop}%
\bibitem [{\citenamefont {Holmes}\ \emph {et~al.}(2013)\citenamefont {Holmes},
  \citenamefont {Ripple},\ and\ \citenamefont
  {Manheimer}}]{holmes_energy-efficient_2013}%
  \BibitemOpen
  \bibfield  {author} {\bibinfo {author} {\bibfnamefont {D.~S.}\ \bibnamefont
  {Holmes}}, \bibinfo {author} {\bibfnamefont {A.~L.}\ \bibnamefont {Ripple}},\
  and\ \bibinfo {author} {\bibfnamefont {M.~A.}\ \bibnamefont {Manheimer}},\
  }\bibfield  {title} {\bibinfo {title} {Energy-{Efficient} {Superconducting}
  {Computing}—{Power} {Budgets} and {Requirements}},\ }\href
  {https://doi.org/10.1109/TASC.2013.2244634} {\bibfield  {journal} {\bibinfo
  {journal} {IEEE Transactions on Applied Superconductivity}\ }\textbf
  {\bibinfo {volume} {23}},\ \bibinfo {pages} {1701610} (\bibinfo {year}
  {2013})},\ \bibinfo {note} {conference Name: IEEE Transactions on Applied
  Superconductivity}\BibitemShut {NoStop}%
\bibitem [{\citenamefont {Toomey}\ \emph {et~al.}(2019)\citenamefont {Toomey},
  \citenamefont {Segall},\ and\ \citenamefont {Berggren}}]{toomey_design_2019}%
  \BibitemOpen
  \bibfield  {author} {\bibinfo {author} {\bibfnamefont {E.}~\bibnamefont
  {Toomey}}, \bibinfo {author} {\bibfnamefont {K.}~\bibnamefont {Segall}},\
  and\ \bibinfo {author} {\bibfnamefont {K.~K.}\ \bibnamefont {Berggren}},\
  }\bibfield  {title} {\bibinfo {title} {Design of a {Power} {Efficient}
  {Artificial} {Neuron} {Using} {Superconducting} {Nanowires}},\ }\bibfield
  {journal} {\bibinfo  {journal} {Frontiers in Neuroscience}\ }\textbf
  {\bibinfo {volume} {13}},\ \href {https://doi.org/10.3389/fnins.2019.00933}
  {10.3389/fnins.2019.00933} (\bibinfo {year} {2019})\BibitemShut {NoStop}%
\bibitem [{\citenamefont {Goteti}\ and\ \citenamefont
  {Dynes}(2021)}]{goteti_superconducting_2021}%
  \BibitemOpen
  \bibfield  {author} {\bibinfo {author} {\bibfnamefont {U.~S.}\ \bibnamefont
  {Goteti}}\ and\ \bibinfo {author} {\bibfnamefont {R.~C.}\ \bibnamefont
  {Dynes}},\ }\bibfield  {title} {\bibinfo {title} {Superconducting neural
  networks with disordered {Josephson} junction array synaptic networks and
  leaky integrate-and-fire loop neurons},\ }\href
  {https://doi.org/10.1063/5.0027997} {\bibfield  {journal} {\bibinfo
  {journal} {Journal of Applied Physics}\ }\textbf {\bibinfo {volume} {129}},\
  \bibinfo {pages} {073901} (\bibinfo {year} {2021})},\ \bibinfo {note}
  {publisher: American Institute of Physics}\BibitemShut {NoStop}%
\bibitem [{\citenamefont {Schneider}\ and\ \citenamefont
  {Segall}(2020)}]{schneider_fan-out_2020}%
  \BibitemOpen
  \bibfield  {author} {\bibinfo {author} {\bibfnamefont {M.~L.}\ \bibnamefont
  {Schneider}}\ and\ \bibinfo {author} {\bibfnamefont {K.}~\bibnamefont
  {Segall}},\ }\bibfield  {title} {\bibinfo {title} {Fan-out and fan-in
  properties of superconducting neuromorphic circuits},\ }\href
  {https://doi.org/10.1063/5.0025168} {\bibfield  {journal} {\bibinfo
  {journal} {Journal of Applied Physics}\ }\textbf {\bibinfo {volume} {128}},\
  \bibinfo {pages} {214903} (\bibinfo {year} {2020})},\ \bibinfo {note}
  {publisher: American Institute of Physics}\BibitemShut {NoStop}%
\bibitem [{\citenamefont {Crotty}\ \emph {et~al.}(2010)\citenamefont {Crotty},
  \citenamefont {Schult},\ and\ \citenamefont
  {Segall}}]{crotty_josephson_2010}%
  \BibitemOpen
  \bibfield  {author} {\bibinfo {author} {\bibfnamefont {P.}~\bibnamefont
  {Crotty}}, \bibinfo {author} {\bibfnamefont {D.}~\bibnamefont {Schult}},\
  and\ \bibinfo {author} {\bibfnamefont {K.}~\bibnamefont {Segall}},\
  }\bibfield  {title} {\bibinfo {title} {Josephson junction simulation of
  neurons},\ }\href {https://doi.org/10.1103/PhysRevE.82.011914} {\bibfield
  {journal} {\bibinfo  {journal} {Physical Review E}\ }\textbf {\bibinfo
  {volume} {82}},\ \bibinfo {pages} {011914} (\bibinfo {year} {2010})},\
  \bibinfo {note} {publisher: American Physical Society}\BibitemShut {NoStop}%
\bibitem [{\citenamefont {Segall}\ \emph {et~al.}(2014)\citenamefont {Segall},
  \citenamefont {Guo}, \citenamefont {Crotty}, \citenamefont {Schult},\ and\
  \citenamefont {Miller}}]{segall_phase-flip_2014}%
  \BibitemOpen
  \bibfield  {author} {\bibinfo {author} {\bibfnamefont {K.}~\bibnamefont
  {Segall}}, \bibinfo {author} {\bibfnamefont {S.}~\bibnamefont {Guo}},
  \bibinfo {author} {\bibfnamefont {P.}~\bibnamefont {Crotty}}, \bibinfo
  {author} {\bibfnamefont {D.}~\bibnamefont {Schult}},\ and\ \bibinfo {author}
  {\bibfnamefont {M.}~\bibnamefont {Miller}},\ }\bibfield  {title}
  {{\selectlanguage {english}\bibinfo {title} {Phase-flip bifurcation in a
  coupled {Josephson} junction neuron system}},\ }\href
  {https://doi.org/10.1016/j.physb.2014.07.048} {\bibfield  {journal} {\bibinfo
   {journal} {Physica B: Condensed Matter}\ }\bibinfo {series} {21st {Latin}
  {American} {Symposium} on {Solid} {State} {Physics} - {SLAFES} 2013},\
  \textbf {\bibinfo {volume} {455}},\ \bibinfo {pages} {71} (\bibinfo {year}
  {2014})}\BibitemShut {NoStop}%
\bibitem [{\citenamefont {Cheng}\ \emph {et~al.}(2019)\citenamefont {Cheng},
  \citenamefont {Goteti},\ and\ \citenamefont
  {Hamilton}}]{cheng_superconducting_2019}%
  \BibitemOpen
  \bibfield  {author} {\bibinfo {author} {\bibfnamefont {R.}~\bibnamefont
  {Cheng}}, \bibinfo {author} {\bibfnamefont {U.~S.}\ \bibnamefont {Goteti}},\
  and\ \bibinfo {author} {\bibfnamefont {M.~C.}\ \bibnamefont {Hamilton}},\
  }\bibfield  {title} {\bibinfo {title} {Superconducting {Neuromorphic}
  {Computing} {Using} {Quantum} {Phase}-{Slip} {Junctions}},\ }\href
  {https://doi.org/10.1109/TASC.2019.2892111} {\bibfield  {journal} {\bibinfo
  {journal} {IEEE Transactions on Applied Superconductivity}\ }\textbf
  {\bibinfo {volume} {29}},\ \bibinfo {pages} {1} (\bibinfo {year} {2019})},\
  \bibinfo {note} {conference Name: IEEE Transactions on Applied
  Superconductivity}\BibitemShut {NoStop}%
\bibitem [{\citenamefont {Cheng}\ \emph {et~al.}(2018)\citenamefont {Cheng},
  \citenamefont {Goteti},\ and\ \citenamefont {Hamilton}}]{cheng_spiking_2018}%
  \BibitemOpen
  \bibfield  {author} {\bibinfo {author} {\bibfnamefont {R.}~\bibnamefont
  {Cheng}}, \bibinfo {author} {\bibfnamefont {U.~S.}\ \bibnamefont {Goteti}},\
  and\ \bibinfo {author} {\bibfnamefont {M.~C.}\ \bibnamefont {Hamilton}},\
  }\bibfield  {title} {{\selectlanguage {english}\bibinfo {title} {Spiking
  neuron circuits using superconducting quantum phase-slip junctions}},\ }\href
  {https://doi.org/10.1063/1.5042421} {\bibfield  {journal} {\bibinfo
  {journal} {Journal of Applied Physics}\ }\textbf {\bibinfo {volume} {124}},\
  \bibinfo {pages} {152126} (\bibinfo {year} {2018})}\BibitemShut {NoStop}%
\bibitem [{\citenamefont {Cheng}\ \emph {et~al.}(2021)\citenamefont {Cheng},
  \citenamefont {Goteti}, \citenamefont {Walker}, \citenamefont {Krause},
  \citenamefont {Oeding},\ and\ \citenamefont {Hamilton}}]{cheng_toward_2021}%
  \BibitemOpen
  \bibfield  {author} {\bibinfo {author} {\bibfnamefont {R.}~\bibnamefont
  {Cheng}}, \bibinfo {author} {\bibfnamefont {U.~S.}\ \bibnamefont {Goteti}},
  \bibinfo {author} {\bibfnamefont {H.}~\bibnamefont {Walker}}, \bibinfo
  {author} {\bibfnamefont {K.~M.}\ \bibnamefont {Krause}}, \bibinfo {author}
  {\bibfnamefont {L.}~\bibnamefont {Oeding}},\ and\ \bibinfo {author}
  {\bibfnamefont {M.~C.}\ \bibnamefont {Hamilton}},\ }\bibfield  {title}
  {\bibinfo {title} {Toward {Learning} in {Neuromorphic} {Circuits} {Based} on
  {Quantum} {Phase} {Slip} {Junctions}},\ }\href
  {https://doi.org/10.3389/fnins.2021.765883} {\bibfield  {journal} {\bibinfo
  {journal} {Frontiers in Neuroscience}\ }\textbf {\bibinfo {volume} {15}},\
  \bibinfo {pages} {1470} (\bibinfo {year} {2021})}\BibitemShut {NoStop}%
\bibitem [{\citenamefont {Schneider}\ \emph {et~al.}(2018)\citenamefont
  {Schneider}, \citenamefont {Donnelly}, \citenamefont {Russek}, \citenamefont
  {Baek}, \citenamefont {Pufall}, \citenamefont {Hopkins}, \citenamefont
  {Dresselhaus}, \citenamefont {Benz},\ and\ \citenamefont
  {Rippard}}]{schneider_ultralow_2018}%
  \BibitemOpen
  \bibfield  {author} {\bibinfo {author} {\bibfnamefont {M.~L.}\ \bibnamefont
  {Schneider}}, \bibinfo {author} {\bibfnamefont {C.~A.}\ \bibnamefont
  {Donnelly}}, \bibinfo {author} {\bibfnamefont {S.~E.}\ \bibnamefont
  {Russek}}, \bibinfo {author} {\bibfnamefont {B.}~\bibnamefont {Baek}},
  \bibinfo {author} {\bibfnamefont {M.~R.}\ \bibnamefont {Pufall}}, \bibinfo
  {author} {\bibfnamefont {P.~F.}\ \bibnamefont {Hopkins}}, \bibinfo {author}
  {\bibfnamefont {P.~D.}\ \bibnamefont {Dresselhaus}}, \bibinfo {author}
  {\bibfnamefont {S.~P.}\ \bibnamefont {Benz}},\ and\ \bibinfo {author}
  {\bibfnamefont {W.~H.}\ \bibnamefont {Rippard}},\ }\bibfield  {title}
  {{\selectlanguage {english}\bibinfo {title} {Ultralow power artificial
  synapses using nanotextured magnetic {Josephson} junctions}},\ }\href
  {https://doi.org/10.1126/sciadv.1701329} {\bibfield  {journal} {\bibinfo
  {journal} {Science Advances}\ }\textbf {\bibinfo {volume} {4}},\ \bibinfo
  {pages} {e1701329} (\bibinfo {year} {2018})}\BibitemShut {NoStop}%
\bibitem [{\citenamefont {Shainline}\ \emph {et~al.}(2017)\citenamefont
  {Shainline}, \citenamefont {Buckley}, \citenamefont {Mirin},\ and\
  \citenamefont {Nam}}]{shainline_superconducting_2017}%
  \BibitemOpen
  \bibfield  {author} {\bibinfo {author} {\bibfnamefont {J.~M.}\ \bibnamefont
  {Shainline}}, \bibinfo {author} {\bibfnamefont {S.~M.}\ \bibnamefont
  {Buckley}}, \bibinfo {author} {\bibfnamefont {R.~P.}\ \bibnamefont {Mirin}},\
  and\ \bibinfo {author} {\bibfnamefont {S.~W.}\ \bibnamefont {Nam}},\
  }\bibfield  {title} {{\selectlanguage {english}\bibinfo {title}
  {Superconducting {Optoelectronic} {Circuits} for {Neuromorphic}
  {Computing}}},\ }\href {https://doi.org/10.1103/PhysRevApplied.7.034013}
  {\bibfield  {journal} {\bibinfo  {journal} {Physical Review Applied}\
  }\textbf {\bibinfo {volume} {7}},\ \bibinfo {pages} {034013} (\bibinfo {year}
  {2017})}\BibitemShut {NoStop}%
\bibitem [{\citenamefont {Shainline}\ \emph {et~al.}(2019)\citenamefont
  {Shainline}, \citenamefont {Buckley}, \citenamefont {McCaughan},
  \citenamefont {Chiles}, \citenamefont {Jafari~Salim}, \citenamefont
  {Castellanos-Beltran}, \citenamefont {Donnelly}, \citenamefont {Schneider},
  \citenamefont {Mirin},\ and\ \citenamefont
  {Nam}}]{shainline_superconducting_2019}%
  \BibitemOpen
  \bibfield  {author} {\bibinfo {author} {\bibfnamefont {J.~M.}\ \bibnamefont
  {Shainline}}, \bibinfo {author} {\bibfnamefont {S.~M.}\ \bibnamefont
  {Buckley}}, \bibinfo {author} {\bibfnamefont {A.~N.}\ \bibnamefont
  {McCaughan}}, \bibinfo {author} {\bibfnamefont {J.~T.}\ \bibnamefont
  {Chiles}}, \bibinfo {author} {\bibfnamefont {A.}~\bibnamefont
  {Jafari~Salim}}, \bibinfo {author} {\bibfnamefont {M.}~\bibnamefont
  {Castellanos-Beltran}}, \bibinfo {author} {\bibfnamefont {C.~A.}\
  \bibnamefont {Donnelly}}, \bibinfo {author} {\bibfnamefont {M.~L.}\
  \bibnamefont {Schneider}}, \bibinfo {author} {\bibfnamefont {R.~P.}\
  \bibnamefont {Mirin}},\ and\ \bibinfo {author} {\bibfnamefont {S.~W.}\
  \bibnamefont {Nam}},\ }\bibfield  {title} {\bibinfo {title} {Superconducting
  optoelectronic loop neurons},\ }\href {https://doi.org/10.1063/1.5096403}
  {\bibfield  {journal} {\bibinfo  {journal} {Journal of Applied Physics}\
  }\textbf {\bibinfo {volume} {126}},\ \bibinfo {pages} {044902} (\bibinfo
  {year} {2019})},\ \bibinfo {note} {publisher: American Institute of
  Physics}\BibitemShut {NoStop}%
\bibitem [{\citenamefont {Krizhevsky}\ \emph {et~al.}(2012)\citenamefont
  {Krizhevsky}, \citenamefont {Sutskever},\ and\ \citenamefont
  {Hinton}}]{krizhevsky_imagenet_2012}%
  \BibitemOpen
  \bibfield  {author} {\bibinfo {author} {\bibfnamefont {A.}~\bibnamefont
  {Krizhevsky}}, \bibinfo {author} {\bibfnamefont {I.}~\bibnamefont
  {Sutskever}},\ and\ \bibinfo {author} {\bibfnamefont {G.~E.}\ \bibnamefont
  {Hinton}},\ }\bibfield  {title} {\bibinfo {title} {{ImageNet}
  {Classification} with {Deep} {Convolutional} {Neural} {Networks}},\ }in\
  \href
  {https://proceedings.neurips.cc/paper/2012/hash/c399862d3b9d6b76c8436e924a68c45b-Abstract.html}
  {\emph {\bibinfo {booktitle} {Advances in {Neural} {Information} {Processing}
  {Systems}}}},\ Vol.~\bibinfo {volume} {25}\ (\bibinfo  {publisher} {Curran
  Associates, Inc.},\ \bibinfo {year} {2012})\BibitemShut {NoStop}%
\bibitem [{\citenamefont {Ye}\ and\ \citenamefont
  {Li}(2021)}]{ye_quantifying_2021}%
  \BibitemOpen
  \bibfield  {author} {\bibinfo {author} {\bibfnamefont {L.}~\bibnamefont
  {Ye}}\ and\ \bibinfo {author} {\bibfnamefont {C.}~\bibnamefont {Li}},\
  }\bibfield  {title} {\bibinfo {title} {Quantifying the {Landscape} of
  {Decision} {Making} {From} {Spiking} {Neural} {Networks}},\ }\href
  {https://doi.org/10.3389/fncom.2021.740601} {\bibfield  {journal} {\bibinfo
  {journal} {Frontiers in Computational Neuroscience}\ }\textbf {\bibinfo
  {volume} {15}},\ \bibinfo {pages} {98} (\bibinfo {year} {2021})}\BibitemShut
  {NoStop}%
\bibitem [{\citenamefont {Tapson}\ \emph {et~al.}(2013)\citenamefont {Tapson},
  \citenamefont {Cohen}, \citenamefont {Afshar}, \citenamefont {Stiefel},
  \citenamefont {Buskila}, \citenamefont {Hamilton},\ and\ \citenamefont {van
  Schaik}}]{tapson_synthesis_2013}%
  \BibitemOpen
  \bibfield  {author} {\bibinfo {author} {\bibfnamefont {J.}~\bibnamefont
  {Tapson}}, \bibinfo {author} {\bibfnamefont {G.}~\bibnamefont {Cohen}},
  \bibinfo {author} {\bibfnamefont {S.}~\bibnamefont {Afshar}}, \bibinfo
  {author} {\bibfnamefont {K.}~\bibnamefont {Stiefel}}, \bibinfo {author}
  {\bibfnamefont {Y.}~\bibnamefont {Buskila}}, \bibinfo {author} {\bibfnamefont
  {T.}~\bibnamefont {Hamilton}},\ and\ \bibinfo {author} {\bibfnamefont
  {A.}~\bibnamefont {van Schaik}},\ }\bibfield  {title} {\bibinfo {title}
  {Synthesis of neural networks for spatio-temporal spike pattern recognition
  and processing},\ }\href {https://doi.org/10.3389/fnins.2013.00153}
  {\bibfield  {journal} {\bibinfo  {journal} {Frontiers in Neuroscience}\
  }\textbf {\bibinfo {volume} {7}},\ \bibinfo {pages} {153} (\bibinfo {year}
  {2013})}\BibitemShut {NoStop}%
\bibitem [{\citenamefont {Maass}(2016)}]{maass_energy-efficient_2016}%
  \BibitemOpen
  \bibfield  {author} {\bibinfo {author} {\bibfnamefont {W.}~\bibnamefont
  {Maass}},\ }\bibfield  {title} {{\selectlanguage {english}\bibinfo {title}
  {Energy-efficient neural network chips approach human recognition
  capabilities}},\ }\href {https://doi.org/10.1073/pnas.1614109113} {\bibfield
  {journal} {\bibinfo  {journal} {Proceedings of the National Academy of
  Sciences}\ }\textbf {\bibinfo {volume} {113}},\ \bibinfo {pages} {11387}
  (\bibinfo {year} {2016})},\ \bibinfo {note} {publisher: National Academy of
  Sciences Section: Commentary}\BibitemShut {NoStop}%
\bibitem [{\citenamefont {Berggren}\ \emph {et~al.}(2018)\citenamefont
  {Berggren}, \citenamefont {Zhao}, \citenamefont {Abebe}, \citenamefont
  {Chen}, \citenamefont {Ravindran}, \citenamefont {McCaughan},\ and\
  \citenamefont {Bardin}}]{berggren_superconducting_2018}%
  \BibitemOpen
  \bibfield  {author} {\bibinfo {author} {\bibfnamefont {K.~K.}\ \bibnamefont
  {Berggren}}, \bibinfo {author} {\bibfnamefont {Q.-Y.}\ \bibnamefont {Zhao}},
  \bibinfo {author} {\bibfnamefont {N.}~\bibnamefont {Abebe}}, \bibinfo
  {author} {\bibfnamefont {M.}~\bibnamefont {Chen}}, \bibinfo {author}
  {\bibfnamefont {P.}~\bibnamefont {Ravindran}}, \bibinfo {author}
  {\bibfnamefont {A.}~\bibnamefont {McCaughan}},\ and\ \bibinfo {author}
  {\bibfnamefont {J.~C.}\ \bibnamefont {Bardin}},\ }\bibfield  {title}
  {{\selectlanguage {english}\bibinfo {title} {A superconducting nanowire can
  be modeled by using {SPICE}}},\ }\href
  {https://doi.org/10.1088/1361-6668/aab149} {\bibfield  {journal} {\bibinfo
  {journal} {Superconductor Science and Technology}\ }\textbf {\bibinfo
  {volume} {31}},\ \bibinfo {pages} {055010} (\bibinfo {year} {2018})},\
  \bibinfo {note} {publisher: IOP Publishing}\BibitemShut {NoStop}%
\bibitem [{\citenamefont {Baghdadi}\ \emph {et~al.}(2020)\citenamefont
  {Baghdadi}, \citenamefont {Allmaras}, \citenamefont {Butters}, \citenamefont
  {Dane}, \citenamefont {Iqbal}, \citenamefont {McCaughan}, \citenamefont
  {Toomey}, \citenamefont {Zhao}, \citenamefont {Kozorezov},\ and\
  \citenamefont {Berggren}}]{baghdadi_multilayered_2020}%
  \BibitemOpen
  \bibfield  {author} {\bibinfo {author} {\bibfnamefont {R.}~\bibnamefont
  {Baghdadi}}, \bibinfo {author} {\bibfnamefont {J.~P.}\ \bibnamefont
  {Allmaras}}, \bibinfo {author} {\bibfnamefont {B.~A.}\ \bibnamefont
  {Butters}}, \bibinfo {author} {\bibfnamefont {A.~E.}\ \bibnamefont {Dane}},
  \bibinfo {author} {\bibfnamefont {S.}~\bibnamefont {Iqbal}}, \bibinfo
  {author} {\bibfnamefont {A.~N.}\ \bibnamefont {McCaughan}}, \bibinfo {author}
  {\bibfnamefont {E.~A.}\ \bibnamefont {Toomey}}, \bibinfo {author}
  {\bibfnamefont {Q.-Y.}\ \bibnamefont {Zhao}}, \bibinfo {author}
  {\bibfnamefont {A.~G.}\ \bibnamefont {Kozorezov}},\ and\ \bibinfo {author}
  {\bibfnamefont {K.~K.}\ \bibnamefont {Berggren}},\ }\bibfield  {title}
  {\bibinfo {title} {Multilayered {Heater} {Nanocryotron}: {A}
  {Superconducting}-{Nanowire}-{Based} {Thermal} {Switch}},\ }\href
  {https://doi.org/10.1103/PhysRevApplied.14.054011} {\bibfield  {journal}
  {\bibinfo  {journal} {Physical Review Applied}\ }\textbf {\bibinfo {volume}
  {14}},\ \bibinfo {pages} {054011} (\bibinfo {year} {2020})},\ \bibinfo {note}
  {publisher: American Physical Society}\BibitemShut {NoStop}%
\bibitem [{\citenamefont {Toomey}\ \emph {et~al.}(2018)\citenamefont {Toomey},
  \citenamefont {Zhao}, \citenamefont {McCaughan},\ and\ \citenamefont
  {Berggren}}]{toomey_frequency_2018}%
  \BibitemOpen
  \bibfield  {author} {\bibinfo {author} {\bibfnamefont {E.}~\bibnamefont
  {Toomey}}, \bibinfo {author} {\bibfnamefont {Q.-Y.}\ \bibnamefont {Zhao}},
  \bibinfo {author} {\bibfnamefont {A.~N.}\ \bibnamefont {McCaughan}},\ and\
  \bibinfo {author} {\bibfnamefont {K.~K.}\ \bibnamefont {Berggren}},\
  }\bibfield  {title} {{\selectlanguage {english}\bibinfo {title} {Frequency
  {Pulling} and {Mixing} of {Relaxation} {Oscillations} in {Superconducting}
  {Nanowires}}},\ }\href {https://doi.org/10.1103/PhysRevApplied.9.064021}
  {\bibfield  {journal} {\bibinfo  {journal} {Physical Review Applied}\
  }\textbf {\bibinfo {volume} {9}},\ \bibinfo {pages} {064021} (\bibinfo {year}
  {2018})}\BibitemShut {NoStop}%
\bibitem [{\citenamefont {Castellani}()}]{castellani_design_nodate}%
  \BibitemOpen
  \bibfield  {author} {\bibinfo {author} {\bibfnamefont {M.}~\bibnamefont
  {Castellani}},\ }\emph {\bibinfo {title} {Design of {Superconducting}
  {Nanowire}-{Based} {Neurons} and {Synapses} for {Power}-{Efficient} {Spiking}
  {Neural} {Networks}}},\ \href@noop {} {Ph.D. thesis}\BibitemShut {NoStop}%
\bibitem [{\citenamefont {Hodgkin}\ and\ \citenamefont
  {Huxley}(1952)}]{hodgkin_quantitative_1952}%
  \BibitemOpen
  \bibfield  {author} {\bibinfo {author} {\bibfnamefont {A.~L.}\ \bibnamefont
  {Hodgkin}}\ and\ \bibinfo {author} {\bibfnamefont {A.~F.}\ \bibnamefont
  {Huxley}},\ }\bibfield  {title} {\bibinfo {title} {A quantitative description
  of membrane current and its application to conduction and excitation in
  nerve},\ }\href {https://www.ncbi.nlm.nih.gov/pmc/articles/PMC1392413/}
  {\bibfield  {journal} {\bibinfo  {journal} {The Journal of Physiology}\
  }\textbf {\bibinfo {volume} {117}},\ \bibinfo {pages} {500} (\bibinfo {year}
  {1952})}\BibitemShut {NoStop}%
\bibitem [{\citenamefont {Toomey}()}]{toomey_superconducting_nodate}%
  \BibitemOpen
  \bibfield  {author} {\bibinfo {author} {\bibfnamefont {E.}~\bibnamefont
  {Toomey}},\ }\emph {\bibinfo {title} {Superconducting nanowire electronics
  for alternative computing}},\ \href@noop {} {Ph.D. thesis}\BibitemShut
  {NoStop}%
\bibitem [{\citenamefont {Teeter}\ \emph {et~al.}(2018)\citenamefont {Teeter},
  \citenamefont {Iyer}, \citenamefont {Menon}, \citenamefont {Gouwens},
  \citenamefont {Feng}, \citenamefont {Berg}, \citenamefont {Szafer},
  \citenamefont {Cain}, \citenamefont {Zeng}, \citenamefont {Hawrylycz},
  \citenamefont {Koch},\ and\ \citenamefont
  {Mihalas}}]{teeter_generalized_2018}%
  \BibitemOpen
  \bibfield  {author} {\bibinfo {author} {\bibfnamefont {C.}~\bibnamefont
  {Teeter}}, \bibinfo {author} {\bibfnamefont {R.}~\bibnamefont {Iyer}},
  \bibinfo {author} {\bibfnamefont {V.}~\bibnamefont {Menon}}, \bibinfo
  {author} {\bibfnamefont {N.}~\bibnamefont {Gouwens}}, \bibinfo {author}
  {\bibfnamefont {D.}~\bibnamefont {Feng}}, \bibinfo {author} {\bibfnamefont
  {J.}~\bibnamefont {Berg}}, \bibinfo {author} {\bibfnamefont {A.}~\bibnamefont
  {Szafer}}, \bibinfo {author} {\bibfnamefont {N.}~\bibnamefont {Cain}},
  \bibinfo {author} {\bibfnamefont {H.}~\bibnamefont {Zeng}}, \bibinfo {author}
  {\bibfnamefont {M.}~\bibnamefont {Hawrylycz}}, \bibinfo {author}
  {\bibfnamefont {C.}~\bibnamefont {Koch}},\ and\ \bibinfo {author}
  {\bibfnamefont {S.}~\bibnamefont {Mihalas}},\ }\bibfield  {title}
  {{\selectlanguage {english}\bibinfo {title} {Generalized leaky
  integrate-and-fire models classify multiple neuron types}},\ }\href
  {https://doi.org/10.1038/s41467-017-02717-4} {\bibfield  {journal} {\bibinfo
  {journal} {Nature Communications}\ }\textbf {\bibinfo {volume} {9}},\
  \bibinfo {pages} {709} (\bibinfo {year} {2018})},\ \bibinfo {note}
  {bandiera\_abtest: a Cc\_license\_type: cc\_by Cg\_type: Nature Research
  Journals Number: 1 Primary\_atype: Research Publisher: Nature Publishing
  Group Subject\_term: Computational neuroscience;Computational science
  Subject\_term\_id:
  computational-neuroscience;computational-science}\BibitemShut {NoStop}%
\bibitem [{\citenamefont {Toomey}\ \emph {et~al.}(2020)\citenamefont {Toomey},
  \citenamefont {Segall}, \citenamefont {Castellani}, \citenamefont
  {Colangelo}, \citenamefont {Lynch},\ and\ \citenamefont
  {Berggren}}]{toomey_superconducting_2020}%
  \BibitemOpen
  \bibfield  {author} {\bibinfo {author} {\bibfnamefont {E.}~\bibnamefont
  {Toomey}}, \bibinfo {author} {\bibfnamefont {K.}~\bibnamefont {Segall}},
  \bibinfo {author} {\bibfnamefont {M.}~\bibnamefont {Castellani}}, \bibinfo
  {author} {\bibfnamefont {M.}~\bibnamefont {Colangelo}}, \bibinfo {author}
  {\bibfnamefont {N.}~\bibnamefont {Lynch}},\ and\ \bibinfo {author}
  {\bibfnamefont {K.~K.}\ \bibnamefont {Berggren}},\ }\bibfield  {title}
  {\bibinfo {title} {Superconducting {Nanowire} {Spiking} {Element} for
  {Neural} {Networks}},\ }\href {https://doi.org/10.1021/acs.nanolett.0c03057}
  {\bibfield  {journal} {\bibinfo  {journal} {Nano Letters}\ }\textbf {\bibinfo
  {volume} {20}},\ \bibinfo {pages} {8059} (\bibinfo {year} {2020})},\ \bibinfo
  {note} {publisher: American Chemical Society}\BibitemShut {NoStop}%
\bibitem [{\citenamefont {Virtanen}\ \emph {et~al.}(2020)\citenamefont
  {Virtanen}, \citenamefont {Gommers}, \citenamefont {Oliphant}, \citenamefont
  {Haberland}, \citenamefont {Reddy}, \citenamefont {Cournapeau}, \citenamefont
  {Burovski}, \citenamefont {Peterson}, \citenamefont {Weckesser},
  \citenamefont {Bright}, \citenamefont {{van der Walt}}, \citenamefont
  {Brett}, \citenamefont {Wilson}, \citenamefont {Millman}, \citenamefont
  {Mayorov}, \citenamefont {Nelson}, \citenamefont {Jones}, \citenamefont
  {Kern}, \citenamefont {Larson}, \citenamefont {Carey}, \citenamefont {Polat},
  \citenamefont {Feng}, \citenamefont {Moore}, \citenamefont {{VanderPlas}},
  \citenamefont {Laxalde}, \citenamefont {Perktold}, \citenamefont {Cimrman},
  \citenamefont {Henriksen}, \citenamefont {Quintero}, \citenamefont {Harris},
  \citenamefont {Archibald}, \citenamefont {Ribeiro}, \citenamefont
  {Pedregosa}, \citenamefont {{van Mulbregt}},\ and\ \citenamefont {{SciPy 1.0
  Contributors}}}]{2020SciPy-NMeth}%
  \BibitemOpen
  \bibfield  {author} {\bibinfo {author} {\bibfnamefont {P.}~\bibnamefont
  {Virtanen}}, \bibinfo {author} {\bibfnamefont {R.}~\bibnamefont {Gommers}},
  \bibinfo {author} {\bibfnamefont {T.~E.}\ \bibnamefont {Oliphant}}, \bibinfo
  {author} {\bibfnamefont {M.}~\bibnamefont {Haberland}}, \bibinfo {author}
  {\bibfnamefont {T.}~\bibnamefont {Reddy}}, \bibinfo {author} {\bibfnamefont
  {D.}~\bibnamefont {Cournapeau}}, \bibinfo {author} {\bibfnamefont
  {E.}~\bibnamefont {Burovski}}, \bibinfo {author} {\bibfnamefont
  {P.}~\bibnamefont {Peterson}}, \bibinfo {author} {\bibfnamefont
  {W.}~\bibnamefont {Weckesser}}, \bibinfo {author} {\bibfnamefont
  {J.}~\bibnamefont {Bright}}, \bibinfo {author} {\bibfnamefont {S.~J.}\
  \bibnamefont {{van der Walt}}}, \bibinfo {author} {\bibfnamefont
  {M.}~\bibnamefont {Brett}}, \bibinfo {author} {\bibfnamefont
  {J.}~\bibnamefont {Wilson}}, \bibinfo {author} {\bibfnamefont {K.~J.}\
  \bibnamefont {Millman}}, \bibinfo {author} {\bibfnamefont {N.}~\bibnamefont
  {Mayorov}}, \bibinfo {author} {\bibfnamefont {A.~R.~J.}\ \bibnamefont
  {Nelson}}, \bibinfo {author} {\bibfnamefont {E.}~\bibnamefont {Jones}},
  \bibinfo {author} {\bibfnamefont {R.}~\bibnamefont {Kern}}, \bibinfo {author}
  {\bibfnamefont {E.}~\bibnamefont {Larson}}, \bibinfo {author} {\bibfnamefont
  {C.~J.}\ \bibnamefont {Carey}}, \bibinfo {author} {\bibfnamefont
  {{\.I}.}~\bibnamefont {Polat}}, \bibinfo {author} {\bibfnamefont
  {Y.}~\bibnamefont {Feng}}, \bibinfo {author} {\bibfnamefont {E.~W.}\
  \bibnamefont {Moore}}, \bibinfo {author} {\bibfnamefont {J.}~\bibnamefont
  {{VanderPlas}}}, \bibinfo {author} {\bibfnamefont {D.}~\bibnamefont
  {Laxalde}}, \bibinfo {author} {\bibfnamefont {J.}~\bibnamefont {Perktold}},
  \bibinfo {author} {\bibfnamefont {R.}~\bibnamefont {Cimrman}}, \bibinfo
  {author} {\bibfnamefont {I.}~\bibnamefont {Henriksen}}, \bibinfo {author}
  {\bibfnamefont {E.~A.}\ \bibnamefont {Quintero}}, \bibinfo {author}
  {\bibfnamefont {C.~R.}\ \bibnamefont {Harris}}, \bibinfo {author}
  {\bibfnamefont {A.~M.}\ \bibnamefont {Archibald}}, \bibinfo {author}
  {\bibfnamefont {A.~H.}\ \bibnamefont {Ribeiro}}, \bibinfo {author}
  {\bibfnamefont {F.}~\bibnamefont {Pedregosa}}, \bibinfo {author}
  {\bibfnamefont {P.}~\bibnamefont {{van Mulbregt}}},\ and\ \bibinfo {author}
  {\bibnamefont {{SciPy 1.0 Contributors}}},\ }\bibfield  {title} {\bibinfo
  {title} {{{SciPy} 1.0: Fundamental Algorithms for Scientific Computing in
  Python}},\ }\href {https://doi.org/10.1038/s41592-019-0686-2} {\bibfield
  {journal} {\bibinfo  {journal} {Nature Methods}\ }\textbf {\bibinfo {volume}
  {17}},\ \bibinfo {pages} {261} (\bibinfo {year} {2020})}\BibitemShut
  {NoStop}%
\bibitem [{Note1()}]{Note1}%
  \BibitemOpen
  \bibinfo {note} {Https://github.com/qnngroup/neuron}\BibitemShut {NoStop}%
\bibitem [{\citenamefont {Chou}\ \emph {et~al.}(2018)\citenamefont {Chou},
  \citenamefont {Chung},\ and\ \citenamefont {Lu}}]{chou_algorithmic_2018}%
  \BibitemOpen
  \bibfield  {author} {\bibinfo {author} {\bibfnamefont {C.-N.}\ \bibnamefont
  {Chou}}, \bibinfo {author} {\bibfnamefont {K.-M.}\ \bibnamefont {Chung}},\
  and\ \bibinfo {author} {\bibfnamefont {C.-J.}\ \bibnamefont {Lu}},\
  }\bibfield  {title} {\bibinfo {title} {On the {Algorithmic} {Power} of
  {Spiking} {Neural} {Networks}},\ }\href {http://arxiv.org/abs/1803.10375}
  {\bibfield  {journal} {\bibinfo  {journal} {arXiv:1803.10375 [cs]}\ }
  (\bibinfo {year} {2018})},\ \bibinfo {note} {arXiv: 1803.10375}\BibitemShut
  {NoStop}%
\bibitem [{\citenamefont {Primavera}\ and\ \citenamefont
  {Shainline}(2021)}]{primavera_active_2021}%
  \BibitemOpen
  \bibfield  {author} {\bibinfo {author} {\bibfnamefont {B.~A.}\ \bibnamefont
  {Primavera}}\ and\ \bibinfo {author} {\bibfnamefont {J.~M.}\ \bibnamefont
  {Shainline}},\ }\bibfield  {title} {\bibinfo {title} {An active dendritic
  tree can mitigate fan-in limitations in superconducting neurons},\ }\href
  {http://arxiv.org/abs/2107.05777} {\bibfield  {journal} {\bibinfo  {journal}
  {arXiv:2107.05777 [cs]}\ } (\bibinfo {year} {2021})},\ \bibinfo {note}
  {arXiv: 2107.05777}\BibitemShut {NoStop}%
\bibitem [{\citenamefont {McCaughan}\ \emph {et~al.}(2018)\citenamefont
  {McCaughan}, \citenamefont {Toomey}, \citenamefont {Schneider}, \citenamefont
  {Berggren},\ and\ \citenamefont
  {Nam}}]{mccaughan_kinetic-inductance-based_2018}%
  \BibitemOpen
  \bibfield  {author} {\bibinfo {author} {\bibfnamefont {A.~N.}\ \bibnamefont
  {McCaughan}}, \bibinfo {author} {\bibfnamefont {E.}~\bibnamefont {Toomey}},
  \bibinfo {author} {\bibfnamefont {M.}~\bibnamefont {Schneider}}, \bibinfo
  {author} {\bibfnamefont {K.~K.}\ \bibnamefont {Berggren}},\ and\ \bibinfo
  {author} {\bibfnamefont {S.~W.}\ \bibnamefont {Nam}},\ }\bibfield  {title}
  {\bibinfo {title} {A kinetic-inductance-based superconducting memory element
  with shunting and sub-nanosecond write times},\ }\href
  {https://doi.org/10.1088/1361-6668/aae50d} {\bibfield  {journal} {\bibinfo
  {journal} {Superconductor science \& technology}\ }\textbf {\bibinfo {volume}
  {32}},\ \bibinfo {pages} {10.1088/1361} (\bibinfo {year} {2018})}\BibitemShut
  {NoStop}%
\bibitem [{\citenamefont {McCaughan}\ and\ \citenamefont
  {Berggren}(2014)}]{mccaughan_superconducting-nanowire_2014}%
  \BibitemOpen
  \bibfield  {author} {\bibinfo {author} {\bibfnamefont {A.~N.}\ \bibnamefont
  {McCaughan}}\ and\ \bibinfo {author} {\bibfnamefont {K.~K.}\ \bibnamefont
  {Berggren}},\ }\bibfield  {title} {\bibinfo {title} {A
  {Superconducting}-{Nanowire} {Three}-{Terminal} {Electrothermal} {Device}},\
  }\href {https://doi.org/10.1021/nl502629x} {\bibfield  {journal} {\bibinfo
  {journal} {Nano Letters}\ }\textbf {\bibinfo {volume} {14}},\ \bibinfo
  {pages} {5748} (\bibinfo {year} {2014})},\ \bibinfo {note} {publisher:
  American Chemical Society}\BibitemShut {NoStop}%
\bibitem [{\citenamefont {Xie}\ \emph {et~al.}(2021)\citenamefont {Xie},
  \citenamefont {Chowdhury}, \citenamefont {Zubair}, \citenamefont {Lozano},
  \citenamefont {Lemettinen}, \citenamefont {Colangelo}, \citenamefont
  {Medeiros}, \citenamefont {Charaev}, \citenamefont {Berggren}, \citenamefont
  {Gumann}, \citenamefont {Pfeiffer},\ and\ \citenamefont
  {Palacios}}]{xie_nbn-gated_2021}%
  \BibitemOpen
  \bibfield  {author} {\bibinfo {author} {\bibfnamefont {Q.}~\bibnamefont
  {Xie}}, \bibinfo {author} {\bibfnamefont {N.}~\bibnamefont {Chowdhury}},
  \bibinfo {author} {\bibfnamefont {A.}~\bibnamefont {Zubair}}, \bibinfo
  {author} {\bibfnamefont {M.~S.}\ \bibnamefont {Lozano}}, \bibinfo {author}
  {\bibfnamefont {J.}~\bibnamefont {Lemettinen}}, \bibinfo {author}
  {\bibfnamefont {M.}~\bibnamefont {Colangelo}}, \bibinfo {author}
  {\bibfnamefont {O.}~\bibnamefont {Medeiros}}, \bibinfo {author}
  {\bibfnamefont {I.}~\bibnamefont {Charaev}}, \bibinfo {author} {\bibfnamefont
  {K.~K.}\ \bibnamefont {Berggren}}, \bibinfo {author} {\bibfnamefont
  {P.}~\bibnamefont {Gumann}}, \bibinfo {author} {\bibfnamefont
  {D.}~\bibnamefont {Pfeiffer}},\ and\ \bibinfo {author} {\bibfnamefont
  {T.}~\bibnamefont {Palacios}},\ }\bibfield  {title} {\bibinfo {title}
  {{NbN}-{Gated} {GaN} {Transistor} {Technology} for {Applications} in
  {Quantum} {Computing} {Systems}},\ }in\ \href@noop {} {\emph {\bibinfo
  {booktitle} {2021 {Symposium} on {VLSI} {Technology}}}}\ (\bibinfo {year}
  {2021})\ pp.\ \bibinfo {pages} {1--2},\ \bibinfo {note} {iSSN:
  2158-9682}\BibitemShut {NoStop}%
\bibitem [{\citenamefont {Pasandi}\ \emph {et~al.}(2019)\citenamefont
  {Pasandi}, \citenamefont {Shafaei},\ and\ \citenamefont
  {Pedram}}]{pasandi_sfqmap_2019}%
  \BibitemOpen
  \bibfield  {author} {\bibinfo {author} {\bibfnamefont {G.}~\bibnamefont
  {Pasandi}}, \bibinfo {author} {\bibfnamefont {A.}~\bibnamefont {Shafaei}},\
  and\ \bibinfo {author} {\bibfnamefont {M.}~\bibnamefont {Pedram}},\
  }\bibfield  {title} {\bibinfo {title} {{SFQmap}: {A} {Technology} {Mapping}
  {Tool} for {Single} {Flux} {Quantum} {Logic} {Circuits}},\ }\href
  {http://arxiv.org/abs/1901.00894} {\bibfield  {journal} {\bibinfo  {journal}
  {arXiv:1901.00894 [quant-ph]}\ } (\bibinfo {year} {2019})},\ \bibinfo {note}
  {arXiv: 1901.00894}\BibitemShut {NoStop}%
\bibitem [{\citenamefont {Zhao}\ \emph {et~al.}(2018)\citenamefont {Zhao},
  \citenamefont {Toomey}, \citenamefont {Butters}, \citenamefont {McCaughan},
  \citenamefont {Dane}, \citenamefont {Nam},\ and\ \citenamefont
  {Berggren}}]{zhao_compact_2018}%
  \BibitemOpen
  \bibfield  {author} {\bibinfo {author} {\bibfnamefont {Q.-Y.}\ \bibnamefont
  {Zhao}}, \bibinfo {author} {\bibfnamefont {E.~A.}\ \bibnamefont {Toomey}},
  \bibinfo {author} {\bibfnamefont {B.~A.}\ \bibnamefont {Butters}}, \bibinfo
  {author} {\bibfnamefont {A.~N.}\ \bibnamefont {McCaughan}}, \bibinfo {author}
  {\bibfnamefont {A.~E.}\ \bibnamefont {Dane}}, \bibinfo {author}
  {\bibfnamefont {S.-W.}\ \bibnamefont {Nam}},\ and\ \bibinfo {author}
  {\bibfnamefont {K.~K.}\ \bibnamefont {Berggren}},\ }\bibfield  {title}
  {{\selectlanguage {english}\bibinfo {title} {A compact superconducting
  nanowire memory element operated by nanowire cryotrons}},\ }\href
  {https://doi.org/10.1088/1361-6668/aaa820} {\bibfield  {journal} {\bibinfo
  {journal} {Superconductor Science and Technology}\ }\textbf {\bibinfo
  {volume} {31}},\ \bibinfo {pages} {035009} (\bibinfo {year}
  {2018})}\BibitemShut {NoStop}%
\end{thebibliography}%

\appendix

\section{Circuit Model}

\begin{figure}[h]
    \centering
    \includegraphics[width=\textwidth]{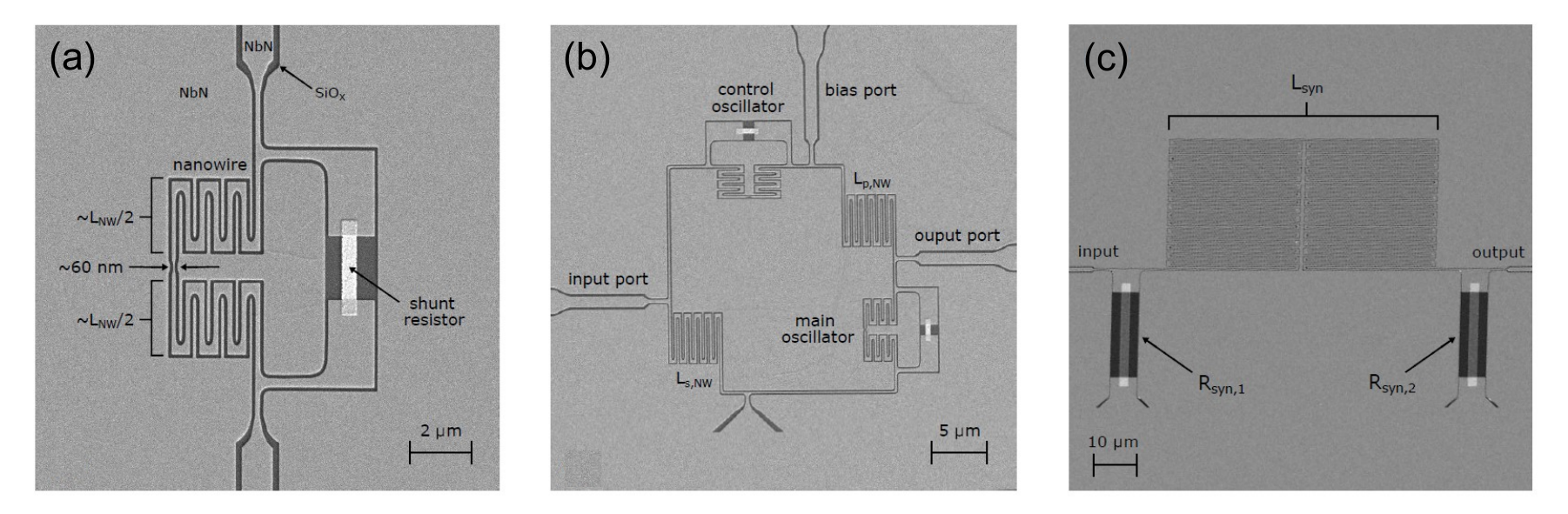}
    \caption{SEM images of fabricated relaxation oscillators (a), nanowire neurons (b), and synaptic integration loop in the synapse (c). Obtained from \cite{castellani_design_nodate}}
    \label{fig:a1}
\end{figure}

\begin{figure}[h]
    \centering
    \includegraphics[width=\textwidth]{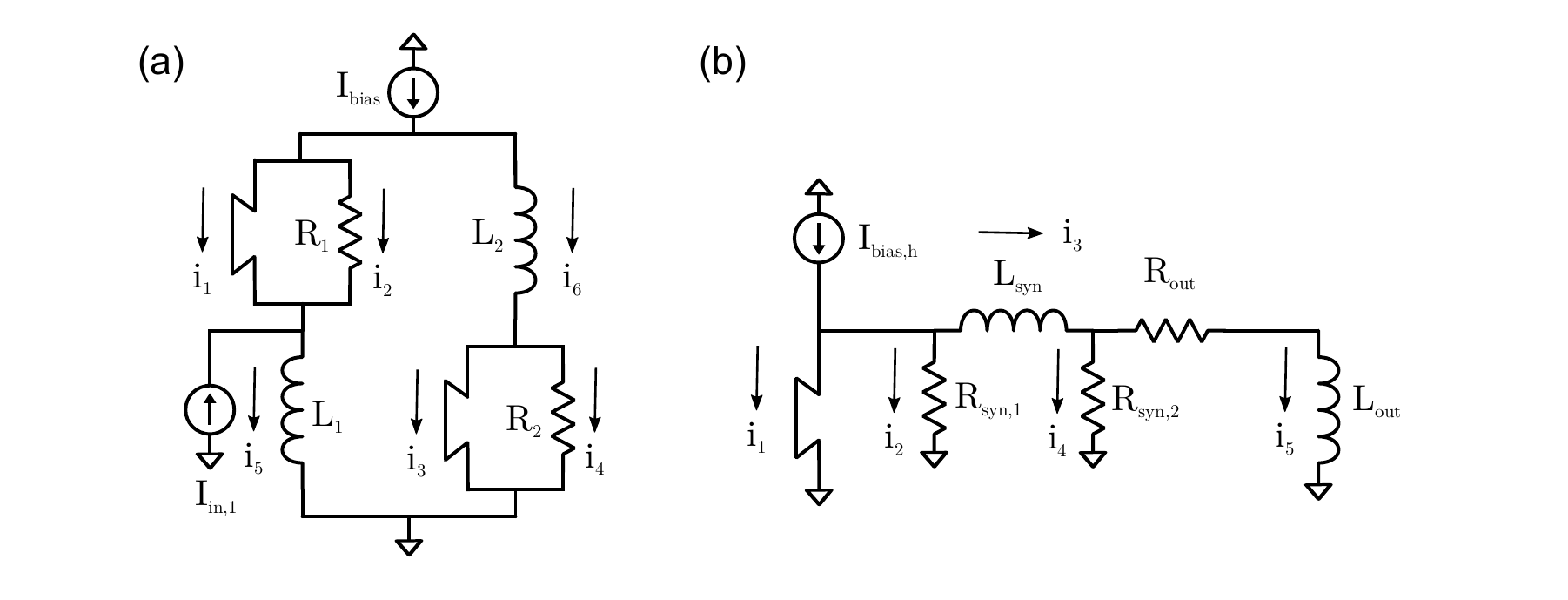}
    \caption{Circuit schematic for the nanowire neuron (a) and the hTron synapse (b) with definitions of currents for a state-description of the circuit in the superconducting state.}
    \label{fig:a2}
\end{figure}

The nanowire neuron is made from two relaxation oscillators. We define the currents $i_1$ through $i_6$ for the nanowire neuron as the currents in each branch of the three loops as in figure \ref{fig:a2}. Applying Kirchhoff's voltage law for the three loops yields the following equations:

\begin{align}
L_{n w}\left(i_{1}\right) \frac{d i_{1}}{d t}+i_{1} R_{h s} n_{1}=i_{2} R_{1} \\
L_{n w}\left(i_{3}\right) \frac{d i_{3}}{d t}+i_{3} R_{h s} n_{2}=i_{4} R_{2} \\
L_{1} \frac{d i_{5}}{d t}+i_{2} R_{1}=L_{2} \frac{d i_{6}}{d t}+i_{3} R_{2}
\end{align}
The hTron synapse is similarly described by the equations from Kirchhoff's voltage law:

\begin{align}
L_{n w, h}\left(i_{1}\right) \frac{d i_{1}}{d t}+i_{1} R_{h s} h=i_{2} R_{s y n, 1} \\
i_{2} R_{s y n, 1}=L_{s y n} \frac{d i_{3}}{d t}+i_{4} R_{s y n, 2} \\
i_{4} R_{s y n, 2}=i_{5} R_{o u t}+L_{o u t} \frac{d i_{5}}{d t}
\end{align}
\noindent
Here $L_{out}$ is taken as $L_2$.
We use state variables $n_1, n_2$ to capture the state of the nanowires in the neuron and $h$ for the channel of the hTron. We modulate the critical current of the hTron according to the following rule:
$$ \text { if } n_{2}=1, \text { then } I_{c, h}{ }^{\prime}=\beta I_{\text {bias }, h} $$
\noindent
Where we define $\beta$ to be a factor such that $0<\beta<1$. This ensures that the hTron channel switches when the nanowire in the main oscillator switches.

\end{document}